\documentclass[twoside,twocolumn,9pt]{article}
\usepackage{extsizes}
\usepackage[super,sort&compress,comma]{natbib} 
\usepackage[version=3]{mhchem}
\usepackage[left=1.5cm, right=1.5cm, top=1.785cm, bottom=2.0cm]{geometry}
\usepackage{balance}
\usepackage{times,mathptmx}
\usepackage{sectsty}
\usepackage{graphicx} 
\usepackage{bm}
\usepackage{latexsym}
\usepackage{epsf}
\usepackage{float}
\usepackage{hyperref}
\usepackage{lastpage}
\usepackage[format=plain,justification=justified,singlelinecheck=false,font={stretch=1.125,small,sf},labelfont=bf,labelsep=space]{caption}
\usepackage{float}
\usepackage{fancyhdr}
\usepackage{fnpos}
\usepackage{revsymb}
\usepackage{verbatim}
\usepackage{amsmath}
\usepackage{subfigure}
\usepackage{caption}
\usepackage{enumerate}
\usepackage{nicefrac}
\usepackage{authblk,etoolbox}
\usepackage{appendix}
\usepackage[hang,flushmargin, symbol]{footmisc}
\usepackage{abstract}
\usepackage{lipsum}
\usepackage{multicol}
\usepackage{hyperref}
\usepackage[usenames,dvipsnames]{xcolor}
\usepackage{braket}
\usepackage[english]{babel}
\addto{\captionsenglish}{%
  
}
\usepackage{mathtools,amssymb,cuted} 
\usepackage{array}
\usepackage{droidsans}
\usepackage{charter}
\usepackage[T1]{fontenc}
\usepackage{setspace}
\usepackage[compact]{titlesec}
\usepackage{hyperref}
\usepackage[normalem]{ulem}

\usepackage{epstopdf}

\definecolor{cream}{RGB}{222,217,201}

\renewcommand\b[1]{{\bf  #1}}
\renewcommand\vec[1]{\boldsymbol{#1}}
\renewcommand\phi{\varphi}
\newcommand\tr{\mathrm{tr}}
\newcommand\del{\nabla}
\newcommand\eps{\epsilon}

\newcommand\dd{\mathrm{d}}

\begin{document}

\pagestyle{fancy}
\thispagestyle{plain}
\fancypagestyle{plain}{

\fancyhead[C]{\includegraphics[width=18.5cm]{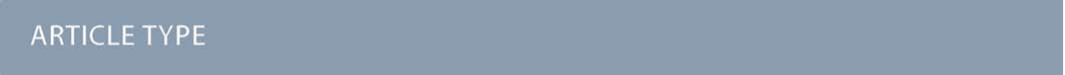}}
\fancyhead[L]{\hspace{0cm}\vspace{1.5cm}\includegraphics[height=30pt]{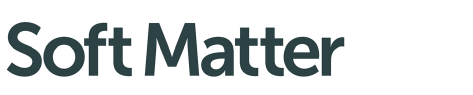}}
\fancyhead[R]{\hspace{0cm}\vspace{1.7cm}\includegraphics[height=55pt]{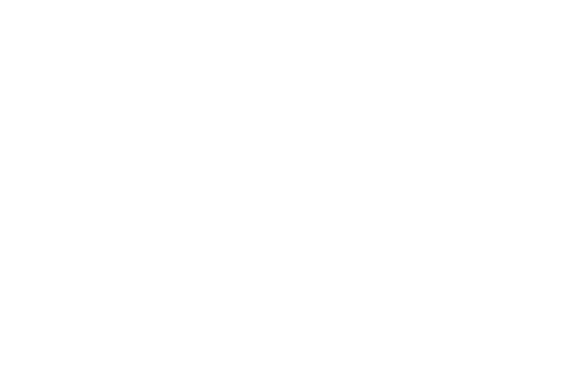}}
\renewcommand{\headrulewidth}{0pt}
}

\makeFNbottom
\makeatletter
\renewcommand\LARGE{\@setfontsize\LARGE{15pt}{17}}
\renewcommand\Large{\@setfontsize\Large{12pt}{14}}
\renewcommand\large{\@setfontsize\large{10pt}{12}}
\renewcommand\footnotesize{\@setfontsize\footnotesize{7pt}{10}}
\makeatother

\renewcommand{\thefootnote}{\fnsymbol{footnote}}
\renewcommand\footnoterule{\vspace*{1pt}%
\color{cream}\hrule width 3.5in height 0.4pt \color{black}\vspace*{5pt}} 
\setcounter{secnumdepth}{5}

\makeatletter 
\renewcommand\@biblabel[1]{#1}            
\renewcommand\@makefntext[1]%
{\noindent\makebox[0pt][r]{\@thefnmark\,}#1}
\makeatother 
\renewcommand{\figurename}{\small{Fig.}~}
\sectionfont{\sffamily\Large}
\subsectionfont{\normalsize}
\subsubsectionfont{\bf}
\setstretch{1.125} 
\setlength{\skip\footins}{0.8cm}
\setlength{\footnotesep}{0.25cm}
\setlength{\jot}{10pt}
\titlespacing*{\section}{0pt}{4pt}{4pt}
\titlespacing*{\subsection}{0pt}{15pt}{1pt}

\fancyfoot{}
\fancyfoot[LO,RE]{\vspace{-7.1pt}\includegraphics[height=9pt]{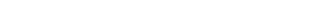}}
\fancyfoot[CO]{\vspace{-7.1pt}\hspace{13.2cm}\includegraphics{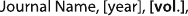}}
\fancyfoot[CE]{\vspace{-7.2pt}\hspace{-14.2cm}\includegraphics{RF.pdf}}
\fancyfoot[RO]{\footnotesize{\sffamily{1--\pageref{LastPage} ~\textbar  \hspace{2pt}\thepage}}}
\fancyfoot[LE]{\footnotesize{\sffamily{\thepage~\textbar\hspace{3.45cm} 1--\pageref{LastPage}}}}
\fancyhead{}
\renewcommand{\headrulewidth}{0pt} 
\renewcommand{\footrulewidth}{0pt}
\setlength{\arrayrulewidth}{1pt}
\setlength{\columnsep}{6.5mm}
\setlength\bibsep{1pt}

\makeatletter 
\newlength{\figrulesep} 
\setlength{\figrulesep}{0.5\textfloatsep} 

\newcommand{\topfigrule}{\vspace*{-1pt}%
\noindent{\color{cream}\rule[-\figrulesep]{\columnwidth}{1.5pt}} }

\newcommand{\botfigrule}{\vspace*{-2pt}%
\noindent{\color{cream}\rule[\figrulesep]{\columnwidth}{1.5pt}} }

\newcommand{\dblfigrule}{\vspace*{-1pt}%
\noindent{\color{cream}\rule[-\figrulesep]{\textwidth}{1.5pt}} }

\makeatother

\twocolumn[
   \begin{@twocolumnfalse}
	\vspace{3cm}
	\sffamily
	\begin{tabular}{m{4.5cm} p{13.5cm} }

		\includegraphics{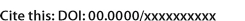} & \noindent\LARGE{\textbf{Shape and size changes of adherent elastic epithelia}} \\
\vspace{0.3cm} & \vspace{0.3cm} \\

 &\noindent\large{Benjamin Loewe$^{a\dag}$, Francesco Serafin$^{b\dag}$, Suraj Shankar$^{c\dag}$, Mark J.~Bowick$^{d}$, and M.~Cristina Marchetti$^{a}$} \\

		\includegraphics{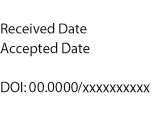} & \noindent\normalsize{Epithelial tissues play a fundamental role in various morphogenetic events during development and early embryogenesis. Although epithelial monolayers are often modeled as two-dimensional (2D) elastic surfaces, they distinguish themselves from conventional thin elastic plates in three important ways- the presence of an apical-basal polarity, spatial variability of cellular thickness, and their nonequilibrium active nature.
		Here, we develop a minimal continuum model of a planar epithelial tissue as an active elastic material that incorporates all these features. We start from a full three-dimensional (3D) description of the tissue and derive an effective 2D model that captures, through the curvature of the apical surface, both the apical-basal asymmetry and the spatial geometry of the tissue. Crucially, variations of active stresses across the apical-basal axis lead to active torques that can drive curvature transitions. By identifying four distinct sources of activity, we find that bulk active stresses arising from actomyosin contractility and growth compete with boundary active tensions due to localized actomyosin cables and lamellipodial activity to generate the various states spanning the morphospace of a planar epithelium.
Our treatment hence unifies 3D shape deformations through the coupled mechanics of apical curvature change and in-plane expansion/contraction of substrate-adhered tissues.
Finally, we discuss the implications of our results for some biologically relevant processes such as tissue folding at the onset of lumen formation.}

\end{tabular}

 \end{@twocolumnfalse} \vspace{0.6cm}

  ]

\renewcommand*\rmdefault{bch}\normalfont\upshape
\rmfamily
\vspace{-1cm}


\footnotetext{\textit{$^{a}$Department of Physics, University of California Santa Barbara, Santa Barbara, California 93106, USA. Email:~baloewe@ucsb.edu}}
\footnotetext{\textit{$^{b}$Department of Physics, University of Michigan, Ann Arbor, Michigan 48109, USA. Email:~fse@umich.edu}}
\footnotetext{\textit{$^{c}$Department of Physics, Harvard University, Cambridge, Massachusetts 02138, USA. Email:~suraj\textunderscore shankar@fas.harvard.edu}}
\footnotetext{\textit{$^{d}$Kavli Institute for Theoretical Physics, University of California, Santa Barbara, California 93106, USA.}}

\footnotetext{\dag These authors contributed equally to this work.}




Living tissues are capable of remarkable deformations and dramatic shape changes key to many developmental processes \cite{lecuit2007orchestrating,nelson2012sculpting}. The diversity of resulting morphogenetic motifs arises from a rich interplay of cell-cell interactions, morphogen gradients and cytoskeletal activity \cite{lecuit2007cell,montell2008morphogenetic}. While the appearance of form along with functionality in living organisms over the course of development involves a plethora of complex biochemical and physiological processes, it has become increasingly clear that mechanics and material approaches offer useful principles to understand the collective organization of cellular matter \cite{trepat2018mesoscale,xi2019material}. In this regard, an important goal of tissue mechanics is to characterize and classify the mechanisms by which thin 2D sheets of cells can fold and deform into 3D shapes. Understanding how shape in biological systems emerges from the spontaneous organization of active processes at the molecular scale remains a grand challenge in biology. It additionally has far reaching implications for the design of self-shaping functional materials \cite{villar2013tissue,ideses2018spontaneous,senoussi2019tunable,morley2019quantitative}.

	A common approach to modeling epithelial tissue mechanics is in analogy with thin sheets of passive elastic or fluid media \cite{gonzalez2012soft,nelson2016buckling}. An important distinction though is that cells actively consume energy to remodel the tissue architecture, thereby allowing the tissue to realize exotic nonequilibrium mechanical properties, ranging from active jammed states \cite{park2015unjamming} to ultradeformable \cite{noll2017active,latorre2018active} and rupture resistant solids \cite{armon2018ultrafast}. In addition, epithelial tissues are intrinsically polarized along the apical-basal axis of the constituent cells, with the basal surface often adhered via a basement membrane to a substrate. This polar asymmetry in conjunction with bulk active stresses, either due to actomyosin contractility \cite{martin2009pulsed,kim2013apical} or growth \cite{dervaux2008morphogenesis,shyer2013villification}, can lead to geometric incompatibilities that shape the tissue \cite{liang2011growth,armon2011geometry}. Importantly, apicobasal polarized active stresses act as torques that compete with both cell-cell and cell-substrate adhesion to pattern differential spatial curvature in the tissue by locally varying the cellular thickness. Previous work has addressed this in the context of the 3D morphology of single epithelial cells \cite{hannezo2014theory}, while continuum modeling on the tissue scale has primarily been restricted to constant thickness shape changes \cite{hannezo2011instabilities,maitra2014activating,berthoumieux2014active,murisic2015discrete,salbreux2017mechanics,mietke2019self} or free monolayers neglecting substrate adhesion \cite{krajnc2015theory,morris2019active}. In the context of wound healing assays and micropatterned tissue cultures, epithelial spreading \cite{serra2012mechanical,kopf2013continuum,brugues2014forces,banerjee2015propagating,blanch2017effective} and dewetting \cite{douezan2012dewetting,ravasio2015regulation,perez2019active} driven by cellular migration and boundary localized active tensions have also been analyzed in the plane without regard to 3D tissue morphology. In the very different context of active suspensions, the wetting properties and shapes of orientationally ordered liquid crystalline drops have been shown to be controlled by an active disjoining pressure and depend on the kinds of topological defects present \cite{joanny2012drop}.

\begin{figure}[]
	\centering
	\includegraphics[width=0.5\textwidth]{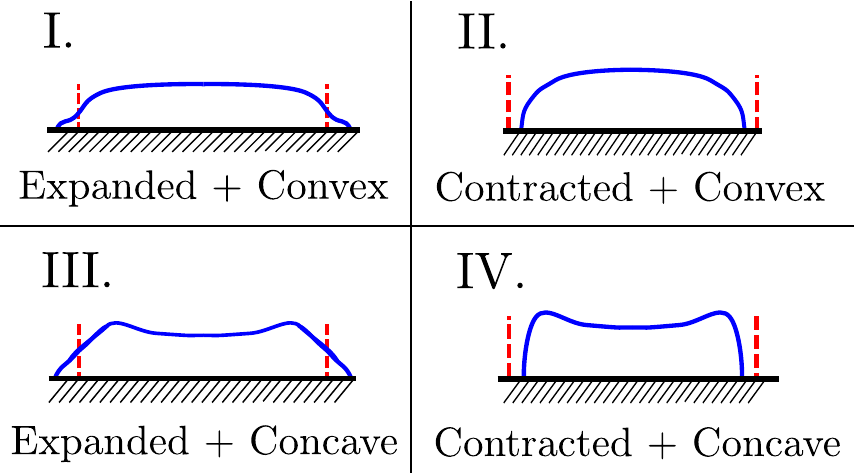}
	\caption{A sketch of the different tissue morphologies that are possible in our model. The red dashed lines mark the extent of the undeformed tissue.}
	\label{fig:shapes}
\end{figure}

	In this paper we derive an effective 2D description for epithelial tissues that accounts for apical-basal polarity, cell-cell interactions and cell-substrate adhesion within an active elastic continuum model. A central feature, apical-basal polarity affects both passive and active sectors of tissue mechanics, allowing active torques in the latter. By exploiting the separation of scales in a thin monolayer, we perform a systematic reduction of the 3D equations of active mechanics to 2D, while retaining the cellular thickness as a dynamical variable. The structure of our equations is consistent with a recently proposed general phenomenological description of active surfaces \cite{salbreux2017mechanics}, with the inclusion of traction forces due to cell-substrate interactions and an explicit derivation of model parameters. By incorporating four distinct sources of cellular activity through nonequilibrium stresses and boundary tensions, our model allows a unified treatment of planar size and apical shape change of substrate-adhered tissues. In particular, we include i) nonequilibrium contributions from bulk contractile stresses due to the apical-medial actomyosin cytoskeleton, ii) extensile stresses generated by cell growth, iii) an apically localized supracellular actomyosin cable that serves as a ``purse-string'', and iv) polarized lamellipodial activity that promotes cell migration at the free boundary of the tissue. The competition of extensile and contractile forces between the boundary and the bulk of the tissue determines its morphology as a function of tissue size and the stiffness of the focal adhesions bound to the substrate. Importantly, differential contractility along the apicobasal axis generates active torques that drive curvature change of the tissue. Working within a simplified 1D setting, we obtain steady-state solutions of our equations that characterize the different possible shapes through the curvature of the apical surface and the in-plane contraction or expansion of the tissue. A cartoon of the shapes predicted by our model is shown in Fig.~\ref{fig:shapes}.

	In Sec.~\ref{sec:model} we introduce the continuum description of an adherent tissue and outline the reduction from 3D to an effective 2D model that incorporates in-plane deformations and variations in the shape of the apical surface. Some details of the tissue parametrization are given in Appendix~\ref{app:strain}. In Sec.~\ref{sec:1dsoln} we examine stationary profiles of the apical surface, tissue deformation and the cellular stress obtained analytically for a one dimensional (1D) geometry corresponding to a tissue layer homogeneous in one of the in-plane directions. In Sec.~\ref{sec:shape} we examine the competition of various active extensile and contractile stresses in controlling tissue shape and identify two transitions, one associated with change in shape of the apical surface, the other with change of in-plane tissue size. Finally we conclude in Sec.~\ref{sec:conclusion} with a brief discussion of the relevance of our model to \emph{in-vitro} experiments of tissue folding and lumen formation.

\section{The model}
\label{sec:model}
\begin{figure}[]
	\centering
    \includegraphics[width=0.5\textwidth]{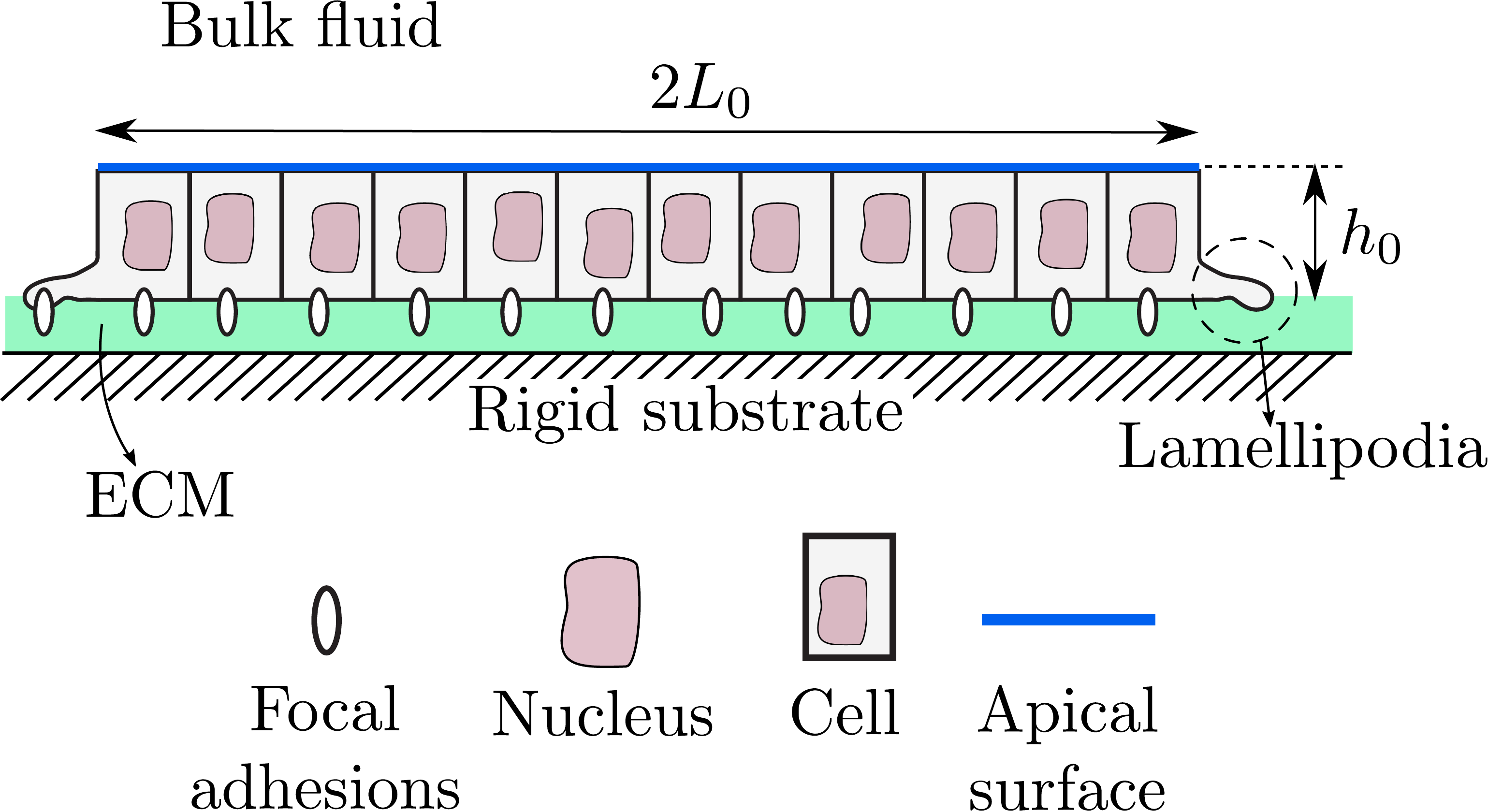}
	\caption{A cartoon of an epithelial monolayer cross-section on a substrate. The apical surface (blue line) is in direct contact with the fluid environment outside and the tissue adheres to the substrate through focal adhesions at the basal surface that bind to the extracellular matrix (ECM), a thin polymeric gel that coats the substrate. The undeformed tissue adopts its rest configuration with linear size $2L_0$ and uniform thickness $h_0$ as shown.}
    \label{fig:cartoon}
\end{figure}

We model an epithelial tissue as a 3D elastic material that is thin in one dimension and adhered to a planar rigid substrate. In the absence of inertia, mechanical equilibrium implies force balance for the 3D stress tensor $\Sigma_{\alpha\beta}$ which gives $\partial_{\beta}\Sigma_{\alpha\beta}=0$. Here and in the following, Greek indices run over all three material coordinates $\{x,y,z\}$, while Roman indices run over only two dimensions, orthogonal to the thin direction, which we take to be $z$.
Writing out the force balance equations, we then have
\begin{align}
	\partial_j\Sigma_{ij}+\partial_{z}\Sigma_{iz}&=0\;,
	\label{eq:fbalance12}\\
	\partial_j\Sigma_{zj}+\partial_z\Sigma_{zz}&=0\;.
	\label{eq:fbalance3}
\end{align}
The rest configuration of the tissue has a linear dimension given by $2L_0$ and thickness $h_0$. In the Lagrangian frame, $z\in[0,h_0]$, where $z=0$ is identified with the basal surface of the tissue and $z=h_0$ is the apical surface (see Fig.~\ref{fig:cartoon}).
	Slowly varying deformations in the $\{x,y\}$ plane then occur on the scale $\sim L_0$, while deformations along $z$ are more rapid, varying on the scale of $h_0$. As $h_0/L_0\ll 1$, Eqs.~\ref{eq:fbalance12},~\ref{eq:fbalance3} generate a heirarchy of stress scales in the bulk of the tissue
\begin{equation}
	\Sigma_{zz}\ll\Sigma_{iz}\ll\Sigma_{ij}\;.\label{eq:heirarchy}
\end{equation}
	This geometric separation of scales underlies the reduction of the 3D model to an effective 2D one, just as for passive shells and plates \cite{ciarlet1997mathematical}. Integrating over $z$, the average 2D stress ($\vec{\sigma}$) and bending moment ($\b{M}$) appear as the first two moments of $\vec{\Sigma}$,
\begin{align}
	\sigma_{ij}&=\int_0^{h_0}\dd z\;\Sigma_{ij}\;,\label{eq:sigmadef}\\
	M_{ij}&=\int_0^{h_0}\dd z\;z~\Sigma_{ij}\;.\label{eq:Mdef}
\end{align}
Averaging Eq.~\ref{eq:fbalance12} over $z$, we obtain an equation for in-plane force balance \cite{banerjee2019continuum}
\begin{equation}
	\partial_j\sigma_{ij}=T_i\;,\label{eq:sigmaeqn}
\end{equation}
where we use $\Sigma_{iz}|_{z=h_0}=0$ as the apical surface is a free surface typically in contact with a fluid, and $T_i\equiv\Sigma_{iz}|_{z=0}$ is the traction force exerted by the tissue on the substrate. Doing the same for the bending moment, we can integrate by parts and use Eqs.~\ref{eq:fbalance12},~\ref{eq:fbalance3} to get the torque balance as
\begin{equation}
	\partial_i\partial_jM_{ij}=f_n\;,\label{eq:Meqn}
\end{equation}
where we employ the symmetry of the stress tensor ($\Sigma_{ij}=\Sigma_{ji}$) and set $f_n\equiv\Sigma_{zz}|_{z=0}-\Sigma_{zz}|_{z=h_0}$ as the net normal force exterted by the tissue. Note that, in the simplest setting within the reduced 2D description, we have three relevant degrees of freedom to capture the total deformation of the tissue, two in-plane displacements and the thickness of the tissue. Eqs.~\ref{eq:sigmaeqn} and~\ref{eq:Meqn} provide a sufficient number of constraints to solve the problem, ensuring it is well-posed. If we wish to retain more degrees of freedom to describe the tissue deformation within an effective 2D model, we can do so by deriving further balance equations for higher moments of $\Sigma_{ij}$ to obtain a consistent description. Given the general setup, we now specialize to the case at hand with a specific constitutive model for the tissue as an active solid.

\subsection{Constitutive relations}
The stress tensor has contributions from both passive elasticity and active stresses ($\vec{\Sigma}=\vec{\Sigma}^{el}+\vec{\Sigma}^{a}$). Assuming a Hookean constitutive law for an isotropic solid, the elastic stress is given by
\begin{equation}
	\Sigma^{el}_{\alpha\beta}=2\bar{\mu}\eps_{\alpha\beta}+\bar{\lambda}\delta_{\alpha\beta}\eps_{\nu\nu}\;,
\end{equation}
where $\eps_{\alpha\beta}$ is the full 3D strain tensor and $\bar{\mu},\bar{\lambda}$ are the 3D Lam{\'e} parameters\footnote{We allow the tissue to be compressible; the incompressible limit can be recovered by taking $\bar{\lambda}\to\infty$.}. The active stress \cite{marchetti2013hydrodynamics,prost2015active} includes two terms, a contractile stress arising from force dipoles exerted by the actomyosin cytoskeleton and an extensile stress accounting for cellular growth. Apicobasal polarity allows us to distinguish the active stress in the $z$ direction versus in the plane, so we separately write
\begin{equation}
	\Sigma_{ij}^a=m\zeta_{\perp}\delta_{ij}+\Omega\delta_{ij}\;,\quad\Sigma^a_{zz}=m\zeta_{\parallel}\;,
\end{equation}
where $m$ is the local density of contractile units, such as phosphorylated myosin motors bound to actin filaments. For simplicity, we take the actomyosin network to be isotropic in the plane with $\zeta_{\perp},\zeta_{\parallel}>0$ controlling the average contractile activity in the plane and along the apicobasal axis respectively. Growth enters as an isotropic extensile pressure ($\Omega<0$) solely in the plane, and we disregard growth in the $z$ direction. An important feature of apicobasal polarity is that the actomyosin cortex is spatially localized near the apical surface. Neglecting any basal myosin for simplicity, we write
\begin{equation}
	m(z)=\dfrac{m_0}{h_0}\dfrac{\sinh(z/\ell)}{\sinh(h_0/\ell)}\;,\label{eq:m}
\end{equation}
where $m_0$ is the concentration of active units at the apical surface and $\ell$ is a localization length. Note the important feature here is the spatial asymmetry of the myosin profile along the apicobasal axis. Such a profile can also be obtained by solving a dynamical equation for the volumetric actomyosin density, $\partial_tm\simeq D(\partial_z^2+\del^2)m-m/\tau$ ($\del^2=\partial_x^2+\partial_y^2$) that combines spatial diffusion ($D$) and turnover of actomyosin units on a time scale $\tau$, while additionally imposing a fixed average $m$ in the cell to capture the mean pool of functional actomyosin whose value is tightly regulated by the cell. For simplicity, we neglect any strain coupling here.
 For $t\gg\tau$ and to lowest order in $\del^2$, $m$ adopts the same profile as in Eq.~\ref{eq:m}, with the localization length $\ell=\sqrt{D\tau}$.

To complete the model reduction to 2D, we follow the standard Kirchoff-Love procedure \cite{ciarlet1997mathematical} and set $\Sigma_{zz}\approx0$ in the tissue interior, as justified by the heirarchy in Eq.~\ref{eq:heirarchy}. This gives,
\begin{equation}
	\eps_{zz}=-\dfrac{\Sigma^a_{zz}+\bar{\lambda}\eps_{kk}}{2\bar{\mu}+\bar{\lambda}}\;.
\end{equation}
Next we set $\Sigma_{iz}\approx 0\implies\eps_{iz}\approx 0$. This permits us to parametrize the $z$ dependence of the strain $\eps_{ij}$ (see Appendix~\ref{app:strain} for derivation) as,
\begin{equation}
	\eps_{ij}(z)=\dfrac{1}{2}\left(\partial_iu_j+\partial_ju_i\right)-\dfrac{h_0}{3}\left(\dfrac{z}{h_0}\right)^3\partial_i\partial_jh\;,\label{eq:epsij}
\end{equation}
where $\b{u}$ is the in-plane displacement and $h$ the local thickness of the deformed tissue. Here we have assumed that the basal surface ($z=0$) does not delaminate from the substrate it is adhered to, and can hence only deform in the plane. We work to linear order in both $\b{u}$ and $h$, as appropriate for small deformations. A fully covariant and nonlinear generalization is easily possible as has been recently done for active surfaces \cite{salbreux2017mechanics,morris2019active}.

Upon using Eqs.~\ref{eq:sigmadef} and~\ref{eq:Mdef}, along with Eq.~\ref{eq:epsij}, we obtain $\vec{\sigma}=\vec{\sigma}^{el}+\vec{\sigma}^c+\vec{\sigma}^g$, where
\begin{gather}
	\sigma_{ij}^{el}=2\mu u_{ij}+\lambda\delta_{ij}\;u_{kk}-\dfrac{\mu h_0}{6}\partial_i\partial_jh-\delta_{ij}\dfrac{\lambda h_0}{12}\del^2h\;,\\
	\sigma_{ij}^c=\left[\zeta_{\perp}-\zeta_{\parallel}\left(\dfrac{\nu}{1-\nu}\right)\right]m_0\ell\;\delta_{ij}\;,\quad\sigma_{ij}^g=h_0\Omega\;\delta_{ij}\;.\label{eq:sigmacg}
\end{gather}
The 2D linearized strain tensor is $u_{ij}=(\partial_iu_j+\partial_ju_i)/2$ and the 2D Lam{\'e} parameters and Poisson ratio are
\begin{equation}
	\mu=\bar{\mu}h_0\;,\quad\lambda=\dfrac{2\bar{\mu}\bar{\lambda}h_0}{(2\bar{\mu}+\bar{\lambda})}\;,\quad\nu=\dfrac{\lambda}{2\mu+\lambda}\;.
\end{equation}
In general we expect $\zeta_{\perp}\gg\zeta_{\parallel}$, as a result, $\sigma_{ij}^c>0$ signalling in-plane contractilility. Also note the presence of $\partial_i\partial_jh$ in the elastic part of the stress tensor, which though unusual, is a natural consequence of apicobasal polarity in passive mechanics, as expected of asymmetric membranes \cite{banerjee2019rolled}. We similarly express the moment tensor as $\b{M}$
\begin{equation}
	M_{ij}=\dfrac{h_0}{2}\sigma_{ij}+\dfrac{h_0}{2}\sigma_{ij}^{c}-\dfrac{\mu h_0^2}{20}\partial_i\partial_j h-\delta_{ij}\dfrac{\lambda h_0^2}{40}\del^2h\;.
	\label{eq:M}
\end{equation}
In all the averages involving $m(z)$ (Eq.~\ref{eq:m}), we assume $\ell\ll h_0$, i.e., the actomyosin density is strongly localized to the apical surface. The first two terms in the moment tensor equation above (Eq.~\ref{eq:M}) also reflect the apico-basal polarity of the tissue.  The first term $h_0\vec{\sigma}/2$ is a ``passive'' contribution that appears because the basal surface is flat and adhered to a substrate, as a result of which the average force taken to act on the mid-plane of the tissue generates a torque on the apical surface. The second term $h_0\vec{\sigma}^c/2$ is an active torque generated by the asymmetric $z$-profile of the actomyosin density (Eq.~\ref{eq:m}). The final two terms in Eq.~\ref{eq:M} are the usual elastic components of the bending moment due to the curvature of the apical surface. Similar active moments have been obtained using inhomogeneous activity profiles in the context of active shells~\cite{berthoumieux2014active}, though for constant thickness surfaces. While recent work \cite{morris2019active} has derived a reduced description of epithelial monolayers keeping track of the tissue thickness, the role of apical-basal polarity was only included in the passive part of the mechanics, and active torques as in Eq.~\ref{eq:M} were missed.

Finally, we specify the constitutive equation for the traction ($\b{T}$) and normal forces ($f_n$) to complete the model description. Assuming the substrate is rigid, we use a viscoelastic model to capture the deformation and turnover of the focal adhesions attached to the substrate. In addition, we introduce an internal in-plane polarization $\b{p}$ that directs individual cell motion. Combining the two, we have \cite{banerjee2015propagating,banerjee2019continuum}
\begin{equation}
	\b{T}=Y_s\b{u}+\Gamma_{\perp}\partial_t\b{u}-f\b{p}\;,\label{eq:T}
\end{equation}
where $Y_s$ is the stiffness of the focal adhesion complexes and $\Gamma_{\perp}$ is an effective friction with the substrate. The active propulsion force $f\b{p}$ accounts for cellular crawling and migration due to actin treadmilling within lamellipodia. In confluent epithelia, the polarization $\b{p}$ is appreciable only near the boundary of the colony \cite{serra2012mechanical,banerjee2015propagating,blanch2017effective}. Following Ref.~\cite{blanch2017effective}, we model the polarization quasi-statically, assuming the tissue is unpolarized in the bulk, and the polarization points along the outward normal at the tissue boundary. Writing to linear order $\partial_t\b{p}=-a\;\b{p}+K\del^2\b{p}$, with $a$ a decay rate and $K$ an elastic constant, we neglect any strain coupling and set $\partial_t\b{p}\approx\b{0}$ to get
\begin{equation}
	\b{p}=\ell_p^2\del^2\b{p}\;,\label{eq:peqn}
\end{equation}
where $\b{p}\cdot\hat{\vec{\nu}}=1$ along the tissue boundary ($\hat{\vec{\nu}}$ is the outward normal).
The localization length $\ell_p=\sqrt{K/a}$ controls the penetration of the polarization into the bulk of the tissue. We assume that the propulsive force is the dominant contribution from polarization, though an active stress $\sim\zeta'\b{p}\b{p}$ \cite{marchetti2013hydrodynamics}, which we neglect, is also generally present. Given the edge localized profile of $\b{p}$, this term also effectively contributes to a boundary stress, albeit one which scales differently with the boundary curvature compared to a line tension (see Sec.~\ref{subsec:bcs}).

The normal force on the tissue has a similar constitutive equation, combining an effective friction ($\Gamma_{\parallel}$) and an apical surface tension ($\gamma$) to give
\begin{equation}
	f_n=\Gamma_{\parallel}\partial_th-\gamma\del^2h\;.\label{eq:fn}
\end{equation}
Note that, here we use the fact that the basal surface does not delaminate from the substrate, hence only vertical distortions of the apical surface, through $h$, contribute to the normal force.
Unlike active membranes with pumps \cite{ramaswamy2000nonequilibrium}, a mean density of actomyosin units at the apical surface ($m_0\neq0$) does not actively induce a finite normal velocity. Instead, bending deformations of the apical surface distort the cytoskeletal network that generates a restoring force $\sim\gamma\del^2h$, through its contractility.

To summarize, using Eqs.~\ref{eq:sigmaeqn} and~\ref{eq:Meqn}, the full set of dynamical equations for the in-plane displacements ($\b{u}$) and the tissue thickness ($h$) are
\begin{gather}
	\Gamma_{\perp}\partial_t\b{u}+Y_s\b{u}=\mu\del^2\b{u}+(\mu+\lambda)\bm{\del}\bm{\del}\cdot\b{u}+\bm{\del}\cdot\left(\vec{\sigma}^c+\vec{\sigma}^g\right)-\dfrac{Bh_0}{12}\bm{\del}\del^2h+f\b{p}\;,\label{eq:ueqn}\\
	\Gamma_{\parallel}\partial_th=\gamma\del^2h-\kappa\del^4h+\dfrac{h_0}{2}\bm{\del}\bm{\del}:\vec{\sigma}^c+\dfrac{h_0}{2}\bm{\del}\cdot\left(Y_s\b{u}+\Gamma_{\perp}\partial_t\b{u}-f\b{p}\right)\;.\label{eq:heqn}
\end{gather}
Here, we have defined $B=2\mu+\lambda$ as the bulk modulus and $\kappa=Bh_0^2/40$ as the bending rigidity of the tissue. Until now, we have addressed three sources of cellular activity, through a contractile stress ($\vec{\sigma}^c$), growth ($\vec{\sigma}^g$) and a propulsive force ($f\b{p}$). The final source of activity appears in the boundary conditions and is discussed below.

\begin{figure}[]
	\centering
	\includegraphics[width=0.5\textwidth]{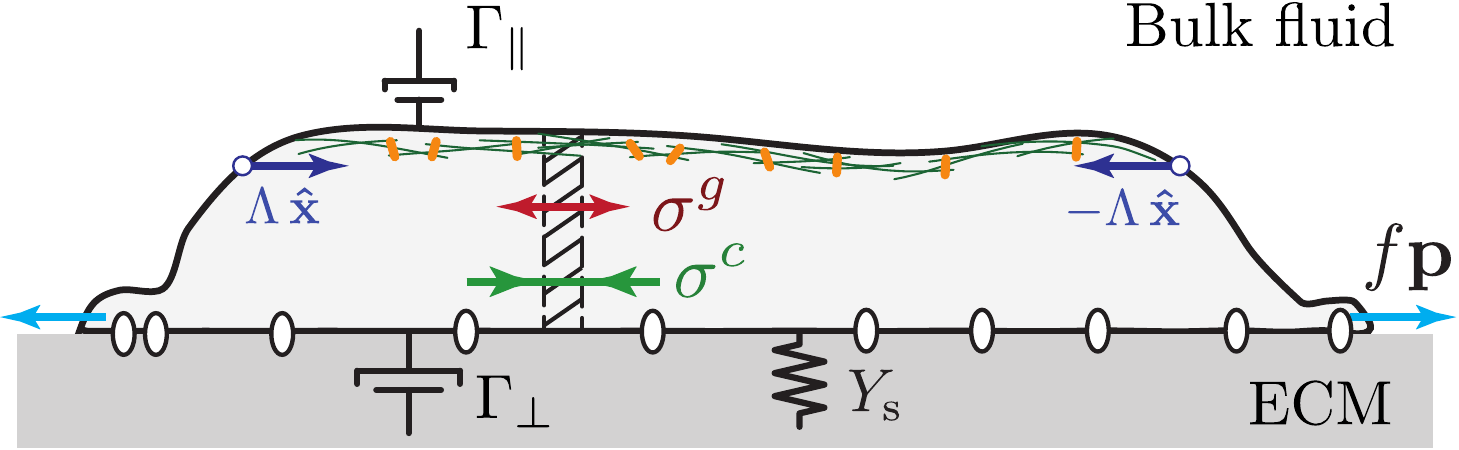}
	\caption{A schematic diagram illustrating the different forces acting on a tissue element.}
	\label{fig:tissue}
\end{figure}
\subsection{Boundary conditions}
\label{subsec:bcs}
Along with the equations of mechanical equilibrium given in Eqs.~\ref{eq:ueqn},~\ref{eq:heqn}, we have to specify the boundary conditions for $\vec{\sigma}$ and $\b{M}$. In doing so we include the presence of a contractile actomyosin cable that is apically localized at the boundary. The assembly of such supracellular structures is known to operate in key morphogenetic events \cite{hutson2003forces,rodriguez2008actomyosin} and wound healing \cite{kiehart1999wound,jacinto2001mechanisms,brugues2014forces}. Boundary actomyosin ``fences'' encircling human stem cell colonies have recently been shown to impact pluripotency as well \cite{narva2017strong}. The simplest way to account for such structures is through a boundary line tension of strength $\tilde{\Lambda}$ localized at the apical surface of the tissue (see Fig.~\ref{fig:tissue}). One can easily show that this also results in an effective boundary torque $\propto h_0\tilde{\Lambda}$. Hence we have
\begin{align}
	\vec{\sigma}\cdot\hat{\vec{\nu}}&=-\tilde{\Lambda}\;C\hat{\vec{\nu}}\;,\\
	\b{M}\cdot\hat{\vec{\nu}}&=-h_0\tilde{\Lambda}\;C\hat{\vec{\nu}}\;,
\end{align}
around the edge of the tissue.
Note that, as expected, the tangential components of both $\vec{\sigma}$ and $\b{M}$ vanish, while the normal components are balanced by the contractile line tension ($\tilde{\Lambda}>0$) along with the boundary curvature ($C$). As before, $\hat{\vec{\nu}}$ is the unit outward normal at the boundary. The various forces acting on the tissue are schematically shown in Fig.~\ref{fig:tissue}. In the following, we will analyze the steady states of the equations we have derived and interpret the solutions in terms of shape changes in the epithelial monolayer.

\section{Stationary Solution in 1D}
\label{sec:1dsoln}
\begin{figure}[]
	\centering
	\includegraphics[width=0.3\textwidth]{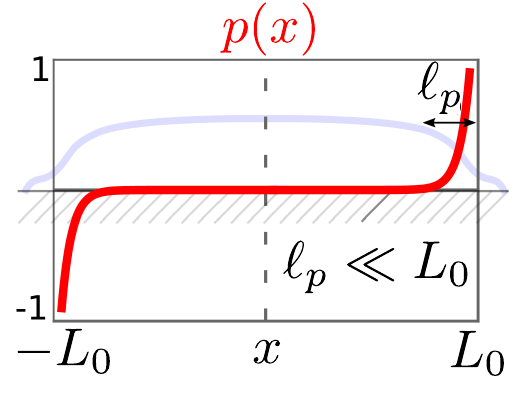}
	\caption{The polarization profile in 1D plotted according to Eq.~\ref{eq:px}.}
	\label{fig:pol}
\end{figure}

For simplicity, we shall work in 1D and assume negligible variation in the $y$-direction. We nonetheless keep a nonzero boundary curvature to represent the effects of the actomyosin cable. Such a simplified description is appropriate in a local 1D strip of a large tissue with curved edges. This turns out to be sufficient to make clear the main features of the model. In Appendix~\ref{app:circle}, we compute the steady-state stress and curvature profile of an axisymmetric 2D tissue in a circular geometry corroborating the validity of the simplified 1D model discussed here. A more detailed treatment of other geometries is left for future work. We choose our coordinate system so that the undeformed tissue has $-L_0\leq x\leq L_0$. Setting $\partial_tu_x=\partial_th=0$, it is convenient to recast Eqs.~\ref{eq:ueqn},~\ref{eq:heqn} in terms of $\sigma_{xx}\equiv\sigma$ and the mean curvature of the apical surface $H=\partial_x^2h$. The equations then read
\begin{align}
	\ell_{\sigma}^2\partial_x^2\sigma&=\sigma-\sigma^c-\sigma^g+\dfrac{Bh_0}{12}H-f\ell_{\sigma}^2\partial_xp\;,\label{eq:sigmaeqn1d}\\
	\ell_{H}^2\partial_x^2H&=H+\dfrac{h_0}{2\gamma}\partial_x^2(\sigma+\sigma^c)\;.\label{eq:Heqn1d}
\end{align}
The stress and curvature relaxation length scales are $\ell_{\sigma}=\sqrt{B/Y_s}$ and $\ell_{H}=\sqrt{\kappa/\gamma}$, respectively, where the bulk modulus $B=2\mu+\lambda$ as before. The active stresses are taken to be spatially constant, $\sigma^c_{xx}\equiv\sigma^c>0$ and $\sigma^g_{xx}\equiv\sigma^g<0$, while the polarization $p(x)$ solves Eq.~\ref{eq:peqn} with $p(\pm L_0)=\pm1$ to give
\begin{equation}
	p(x)=\dfrac{\sinh(x/\ell_p)}{\sinh(L_0/\ell_p)}\;,\label{eq:px}
\end{equation}
which is sketched in Fig.~\ref{fig:pol}. For the boundary conditions, as mentioned earlier, we fix the boundary curvature $C=C_0$ to be a constant and write $\Lambda=\tilde{\Lambda}C_0$ as the effective normal stress at the boundary due to the actomyosin cable. In a circular geometry of size $R$, $C_0=1/R$ and the boundary stress then depends on the tissue size for constant $\tilde{\Lambda}$ (see Appendix~\ref{app:circle}). For generality, we work with $\Lambda$ as an independent parameter, keeping in mind that for certain geometries there could be an implicit tissue size dependence in it. Also note that the anisotropic active stress $\sim \zeta'\b{p}\b{p}$ that we neglect here could potentially contribute a boundary curvature independent term to $\Lambda$ by virtue of the edge localized profile of $\b{p}$. This further justifies our use of $\Lambda$ as an independent parameter.
The full analytical solution of the above equations along with the requisite boundary conditions ($\sigma(\pm L_0)=-\Lambda$, $M(\pm L_0)=-h_0\Lambda$) is not very illuminating. Instead it is instructive to consider the case where surface tension dominates bending elasticity, allowing us to neglect $\ell_{H}^2\partial_x^2H\ll H$ and directly slave the tissue curvature to the stress profile as
\begin{equation}
	H\simeq-\dfrac{h_0}{2\gamma}\partial_x^2\sigma\;,\label{eq:slave}
\end{equation}
where $\sigma^c$ has dropped out as it is a constant. Of course, this approximation will fail close to the tissue boundary where, in particular, the line tension $\Lambda>0$ requires $H(L_0)=h_0(\sigma^c+3\Lambda)/2\kappa>0$ at the edge. Substituting Eq.~\ref{eq:slave} into Eq.~\ref{eq:sigmaeqn1d} we find that, in this limit, apicobasal polarity simply affects the passive mechanics by enhancing the stress relaxation length scale to $L_{\sigma}^2=\ell_\sigma^2+(Bh_0^2/24\gamma)$. So we have
\begin{equation}
	L_{\sigma}^2\partial_x^2\sigma-\sigma=-\left(\sigma^c+\sigma^g+f\ell_{\sigma}^2\partial_xp\right)\;.\label{eq:sig0}
\end{equation}
Upon imposing $\sigma(\pm L_0)=-\Lambda$, we obtain the spatial stress profile in the tissue to be
\begin{align}
	\sigma(x)&=\sigma_a-(\Lambda+\sigma_a)\dfrac{\cosh(x/L_\sigma)}{\cosh(L_0/L_{\sigma})}\nonumber\\
	 &+f\dfrac{\ell_{\sigma}^2\ell_p}{\ell_p^2-L_{\sigma}^2}\left[\dfrac{\cosh(x/\ell_p)}{\sinh(L_0/\ell_p)}-\coth\left(\dfrac{L_0}{\ell_p}\right)\dfrac{\cosh(x/L_\sigma)}{\cosh(L_0/L_\sigma)}\right]\;.
	\label{eq:sigmasoln}
\end{align}
We have combined the two bulk active stresses into $\sigma_a=\sigma^c+\sigma^g$. When actomyosin contractility dominates growth $\sigma_a>0$ and when growth dominates $\sigma_a<0$. Qualitatively similar stress profiles have been obtained using a fluid model for a tissue~\cite{blanch2017effective}.
With these approximations Eq.~\ref{eq:slave} directly gives us the curvature profile of the apical surface as
\begin{align}
	H(x)&=\dfrac{h_0}{2\gamma}\left\{\dfrac{(\Lambda+\sigma_a)}{L_\sigma^2}\dfrac{\cosh(x/L_\sigma)}{\cosh(L_0/L_\sigma)}\right.\nonumber\\
	&\left.-\dfrac{f\;\ell_{\sigma}^2}{\ell_p(\ell_p^2-L_{\sigma}^2)}\left[\dfrac{\cosh(x/\ell_p)}{\sinh(L_0/\ell_p)}-\left(\dfrac{\ell_p}{L_\sigma}\right)^2\coth\left(\dfrac{L_0}{\ell_p}\right)\dfrac{\cosh(x/L_\sigma)}{\cosh(L_0/L_\sigma)}\right]\right\}\;.
	\label{eq:Hsoln}
\end{align}
Similarly, using the steady state in-plane force balance $\partial_x\sigma=Y_su_x-fp$, we find the displacement of the tissue to be
\begin{align}
	u_x(x)&=-\dfrac{(\Lambda+\sigma_a)}{Y_sL_{\sigma}}\dfrac{\sinh(x/L_{\sigma})}{\cosh(L_0/L_{\sigma})}+\dfrac{f}{Y_s}\left\{\dfrac{\sinh(x/\ell_p)}{\sinh(L_0/\ell_p)}\right.\nonumber\\
	&+\left.\dfrac{\ell_{\sigma}^2}{\ell_p^2-L_\sigma^2}\left[\dfrac{\sinh(x/\ell_p)}{\sinh(L_0/\ell_p)}-\dfrac{\ell_p}{L_\sigma}\coth\left(\dfrac{L_0}{\ell_p}\right)\dfrac{\sinh(x/L_\sigma)}{\cosh(L_0/L_\sigma)}\right]\right\}\;.
	\label{eq:usoln}
\end{align}
\section{Active shaping of planar epithelia}
\label{sec:shape}
We now use the above solution (Eqs.~\ref{eq:sigmasoln},~\ref{eq:Hsoln},~\ref{eq:usoln}) to interpret and characterize the morphology of an adhered epithelium. The curvature of the apical surface at the center of the tissue $H(0)$ and the displacement at the edge $u_x(L_0)$ serve as simple ``order parameters'' characterizing the shape of the tissue. Note that, when $H(0)>0$, the apical surface curves up, away from the substrate, adopting an upward concave profile (``valley shaped''), while for $H(0)<0$, we have an upward convex profile (``dome shaped'') for the apical surface of the tissue. Separately, the in-plane displacement of the tissue edge $u_x(L_0)$ tracks the overall expansion ($u_x(L_0)>0$) or contraction ($u_x(L_0)<0$) of the tissue with respect to its undeformed state. Note that while the expanded or contracted state is a steady state of the elastic tissue with no spreading, the tendency towards larger or smaller contact areas in our elastic model is analogous to wetting/dewetting within a fluid model \cite{douezan2012dewetting,perez2019active}. The equations for a fluid tissue are formally identical to those for an elastic tissue, with the flow velocity replacing the displacement field, although the tissue can of course spread.

\subsection{Role of growth, contractility and the actomyosin cable}
We shall first consider the simple case where the polarized motility of the leading cells at the edge of the epithelium is absent, by setting $f=0$.
The competition between adhesion to the substrate and elastic and active stresses creates a spatially inhomogeneous stress profile in the resting tissue sheet. If active stresses are homogeneous, as we consider, the length scale controlling spatial inhomogeneities $\sim L_{\sigma}$ is determined primarily by the relative strength of tissue to focal adhesion elasticity. In this case, the stress profile is monotonic between $x=0$ and $x=L_0$, and symmetric across $x=0$.
As expressed in Eq.~\ref{eq:slave}, spatial inhomogeneities in the stress alone result in a nonvanishing curvature of the apical surface with $H\propto-\partial_x^2\sigma$ (a homogeneous stress profile always yields a flat surface). 

It is possible to obtain a change in the sign of $H(x)$ even in the absence of line tension from the actomyosin cable ($\Lambda=0$), simply from the competition between contractile and extensile uniform active stresses, as  in this case $H(x)\propto\sigma_a$, with $\sigma_a=\sigma^c+\sigma^g$. Additionally, $u_x(L_0)\propto-\sigma_a\sinh(L_0/L_{\sigma})$ (from Eq.~\ref{eq:usoln}), hence the sign of $\sigma_a$ controls the behavior as follows:
\begin{itemize}
	\item If contractile stresses exceed extensile ones ($\sigma_a>0$), then the tissue stress is everywhere contractile (positive, like a negative pressure) and maximum at the center of the tissue. Correspondingly, $H(x)>0$, i.e., the apical surface is shaped like a valley, as one would physically expect from a decrease in internal pressure, and $u(L_0)<0$, i.e., the tissue is contracted (see Fig.~\ref{fig:shapes}, image IV).
	\item If extensile stresses exceed contractile ones ($\sigma_a<0$), then the tissue stress is everywhere extensile (negative, like a positive pressure) and maximum at the edges of the tissue. Correspondingly, $H(x)<0$, i.e., the apical surface is shaped like a dome, as one would physically expect from an increase in internal pressure, and $u(L_0)>0$, i.e., the tissue expands on the substrate (see Fig.~\ref{fig:shapes}, image I).
\end{itemize}
Reinstating the actomyosin cable tension $\Lambda>0$ makes the stress profile more negative, with now contractile behavior (and $H(x)>0$) arising when $\Lambda+\sigma_a>0$ and extensile (and $H(x)<0$) arising when $\Lambda+\sigma_a<0$, upon neglecting the irrelevant constant $\sigma_a$ offset (see Eq.~\ref{eq:sigmasoln}). This can be reformulated in terms of the value of $\Lambda$ required for the two different shapes, with
\begin{equation}
	\Lambda_c=-\sigma_a=-(\sigma^g+\sigma^c)\;.\label{eq:Lambdac1}
\end{equation}
\begin{figure}[]
    \centering
    \includegraphics[width=0.5\textwidth]{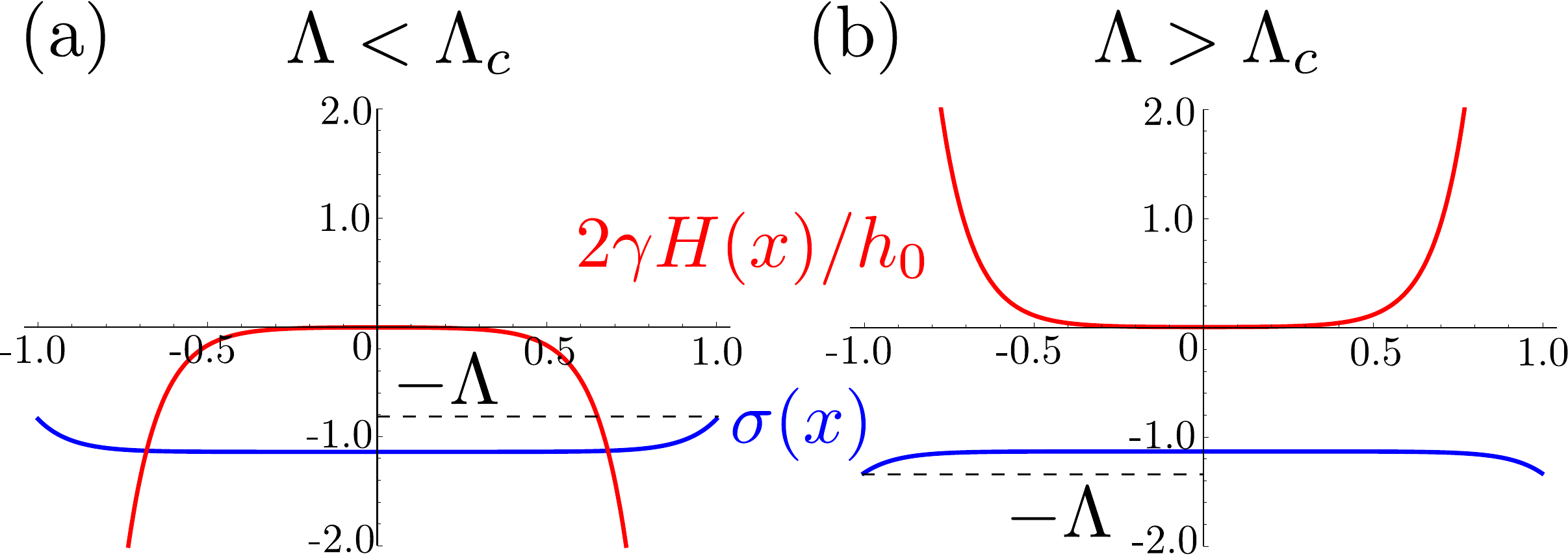}
	\caption{The stress and curvature distribution across the tissue with a bulk active stress ($\sigma_a$) and a boundary tension ($\Lambda$). For dominantly extensile stresses ($\sigma_a<0$), there is a finite threshold $\Lambda_c=-\sigma_a$ (Eq.~\ref{eq:Lambdac1}) for the tension beyond which the curvature of the apical surface changes sign. The spatial profiles of $\sigma(x)$ and $H(x)$ are plotted here in units where $B=1$ and $L_0=1$, neglecting $\ell_H$, for (a) $\Lambda<\Lambda_c$ and (b) $\Lambda>\Lambda_c$. Note that the stress at the boundary of the tissue is given by $-\Lambda$ as required by the boundary condition.}
    \label{fig:siga}
\end{figure}
So a curvature transition can only occur if $|\sigma^g|>|\sigma^c|$ (as $\sigma^g<0$, being extensile, and we must have $\Lambda_c>0$). 
The displacement field of the tissue at the boundary is in this case is $u_x(L_0)\propto-(\Lambda+\sigma_a)\sinh(L_0/L_{\sigma})$ (Eq.~\ref{eq:usoln}). As a result the tissue also undergoes an elastic size transition at a value of $\Lambda$ that coincides with the change in apical surface curvature. Hence, we find
\begin{itemize}
	\item $\Lambda>\Lambda_c$: contractile behavior with the stress peaked at the middle of the contracted tissue leading to a valley-shaped apical surface and contracted tissue (see Fig.~\ref{fig:shapes}, image IV).
	\item $\Lambda<\Lambda_c$: extensile behavior with the stress peaked at the edges of the expanded tissue leading to a dome-shaped apical surface and expanded tissue (see Fig.~\ref{fig:shapes}, image I).
\end{itemize}
In short, when growth dominates contractility ($\sigma_a<0$), an increase in the tension of the actomyosin cable beyond the threshold $\Lambda_c$ causes the tissue to transition from dome-shaped to valley-shaped. The spatial profile of the stress and curvature are plotted in Fig.~\ref{fig:siga}.
For a large tissue ($L_0\gg L_\sigma$), one always has
\begin{equation}
	\sigma(0)\simeq\sigma_a\;,\quad H(0)\simeq 0\;.\label{eq:sigma0}
\end{equation}
Since $\sigma(L_0)=-\Lambda<0$, and the stress is monotonic, one then has $\sigma(x)<0$ everywhere (see Fig.~\ref{fig:siga}) if $\sigma_a<0$ (required for $\Lambda_c$ to exist). Hence the stresses are always extensile, but can still be maximum in the middle or at the edges, with a corresponding change in the sign of $H(x)$ depending on the value of $\Lambda$ relative to $\Lambda_c$. In this case the value of $H(0)$ alone, being exponentially small in a large tissue, does not provide a good criterion for the sign of the curvature, while the full curvature profile is still meaningful. On the other hand, when contractile active stresses dominate growth, the apical surface always adopts a valley like profile and the tissue contracts, no matter the strength of the actomyosin line tension.

\subsection{Role of polarized cell motility}
Now for $f\neq 0$, we have an additional length scale $\ell_p$ in the problem that can compete against $L_{\sigma}$, allowing both the stress and curvature profile to become nonmonotonic on $0\leq x\leq L_0$.
This yields two distinct $\Lambda$ thresholds, one for change in curvature of the apical surface and the other for tissue size change, allowing for the four tissue shapes shown schematically in Fig.~\ref{fig:shapes}.

\begin{figure}[]
    \centering
    \includegraphics[width=0.5\textwidth]{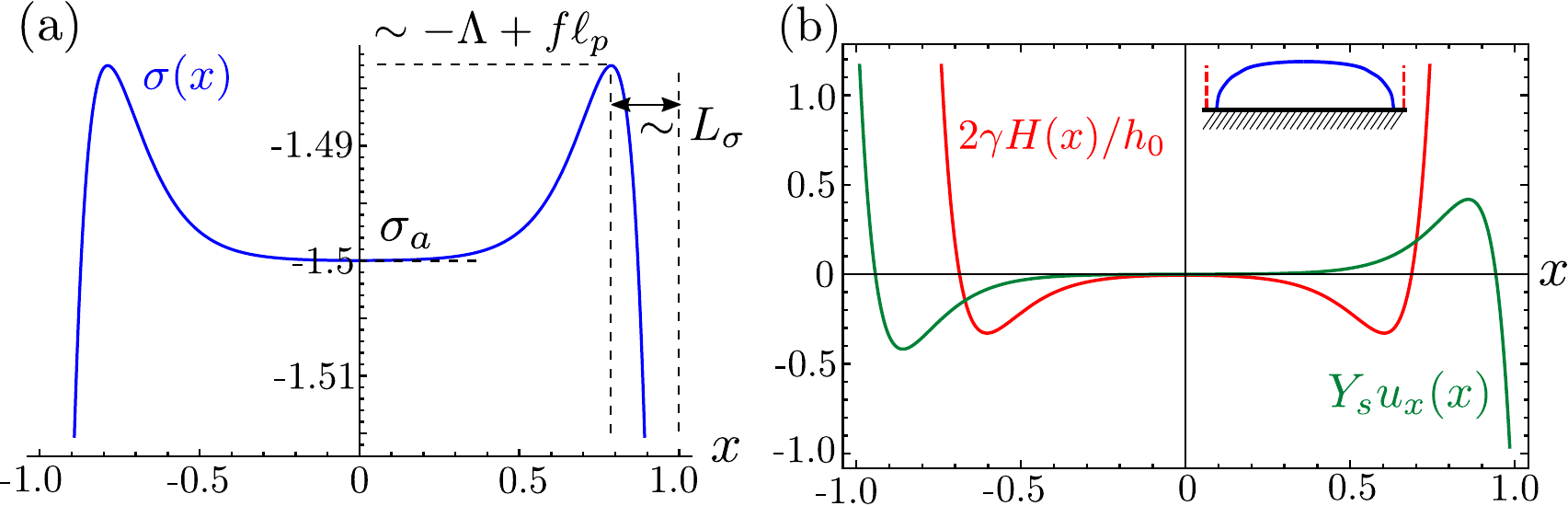}
	\caption{A representative plot showing the nonmonotonic spatial variation of (a) the stress $\sigma(x)$ along with (b) the curvature $H(x)$ and the displacement $u_x(x)$. Here we have taken $\ell_p/\ell_\sigma=0.7$ and $\ell_\sigma/L_0=0.1$, along with $\sigma_a=-1.5<0$ (in units with $B=1$). $\Lambda$ is chosen to lie between $\Lambda_d$ and $\Lambda_c$ ($f\neq 0$). As $L_\sigma\sim\ell_\sigma\ll L_0$, the stress at the center of the tissue is $\sim\sigma_a$ (Eq.~\ref{eq:sigma0}), while self-propulsion at the tissue edge generates an extensile stress $\sim f\ell_p$ in excess of the boundary tension $\Lambda$. The resulting stress peak localized on a scale $\sim L_\sigma$ near the boundary leads to the nonmonotonic behaviour of both $H(x)$ and $u_x(x)$ shown in (b). As shown schematically at the top of plot (b), this corresponds to the case when the tissue has contracted at its edge and has a convex shape in the interior.}
    \label{fig:nonmon}
\end{figure}

Before we address the fully general case, let us first switch off all bulk sources of activity ($\sigma_a=0$). While $\sigma(L_0)=-\Lambda<0$ still, the stress at the center of the tissue can change sign and so can its curvature ($\partial_x^2\sigma$). The propulsive force at the edge of the tissue enhances the stress in a region of width controlled roughly by $\max(L_{\sigma},\ell_p)\ll L_0$, leading (for a sufficiently large $f$) to a positve stress peak $\sim f\ell_p$ localized near the boundary (see Fig.~\ref{fig:nonmon}a). The physics in this case is akin to that of a stretched rubber band attached to a rigid surface, with the pre-stretch combining the net competition between the contractile ring and the propulsive force.

Putting back the bulk active stress $\sigma_a$, the nonmonotonic stress profile persists, which in turn allows for two distinct transition thresholds for the apical curvature change and for elastic size change. Setting $x=0$ in Eq.~\ref{eq:Hsoln}, we have
\begin{gather}
	H(0)=\dfrac{h_0}{2\gamma\cosh(L_0/L_\sigma)L_{\sigma}^2}\left[\sigma_a-(fL_c-\Lambda)\right]\;,\label{eq:H0}\\
	L_c=\dfrac{\ell_{\sigma}^2}{\ell_p\left(\ell_p^2-L_{\sigma}^2\right)}\left[\dfrac{L_{\sigma}^2\cosh(L_0/L_\sigma)-\ell_p^2\cosh(L_0/\ell_p)}{\sinh(L_0/\ell_p)}\right]\;.
\end{gather}
The length scale $L_c$ represents the effective region over which the propulsive force accumulates stress and affects the apical surface curvature. Using the fact that $x^2\cosh(1/x)$ is a positive and monotonically decreasing function until its minimum at $x\approx0.48$, one can show that $L_c>0$ for $\ell_{p},L_{\sigma}\lesssim0.48 L_0$. Note that $L_c$ also remains smooth and finite for $\ell_p=L_{\sigma}$, and is hence a legitimate length scale in the physical regime of interest. For $\ell_p\ll L_{\sigma}$ in a large tissue ($L_0\gg L_\sigma,\ell_p$), we have $L_c\simeq\ell_p(\ell_\sigma/L_\sigma)^2\sim\ell_p$ as expected. Interestingly though, for $\ell_p\simeq L_{\sigma}$, we find $L_c\simeq L_0(\ell_\sigma^2/2L_{\sigma}\ell_p)$ and when $\ell_p\gg L_{\sigma}$, $L_c$ grows exponentially large in the tissue size. This dramatic enhancement of the region of influence of the polarized motility for $\ell_p\gtrsim L_\sigma$ through the tissue and focal adhesion elasticity is reminiscent of similar collective force transmission seen in expanding monolayers \cite{trepat2009physical}.

From Eq.~\ref{eq:H0}, we immediately find that $H(0)$ changes sign at a threshold actomyosin cable tension,
\begin{equation}
	\Lambda_c=fL_c-\sigma_a\;.\label{eq:Lambdac2}
\end{equation}
As expected, the propulsive force increases the threshold for the tissue shape  transition. So for
\begin{itemize}
	\item $\Lambda>\Lambda_c$: tissue adopts a valley-shaped apical surface.
	\item $\Lambda<\Lambda_c$: tissue adopts a dome-shaped apical surface.
\end{itemize}
Recall that $L_c$ can be very large in a large tissue when $\ell_p\gtrsim L_\sigma$, which suggests that such a shape transition can only be realistically observed in smaller tissues or when the polarization is very strongly localized ($\ell_p\ll L_\sigma$).
Of course this only refers to the curvature near the center of the tissue. The nonmonotonic spatial profile of the stress and curvature implies that the shape of the apical surface can also change close to the boundary. A representative plot of such a curvature profile is shown in Fig.~\ref{fig:nonmon}.

\begin{figure}[]
	\centering
	\includegraphics[width=0.40\textwidth]{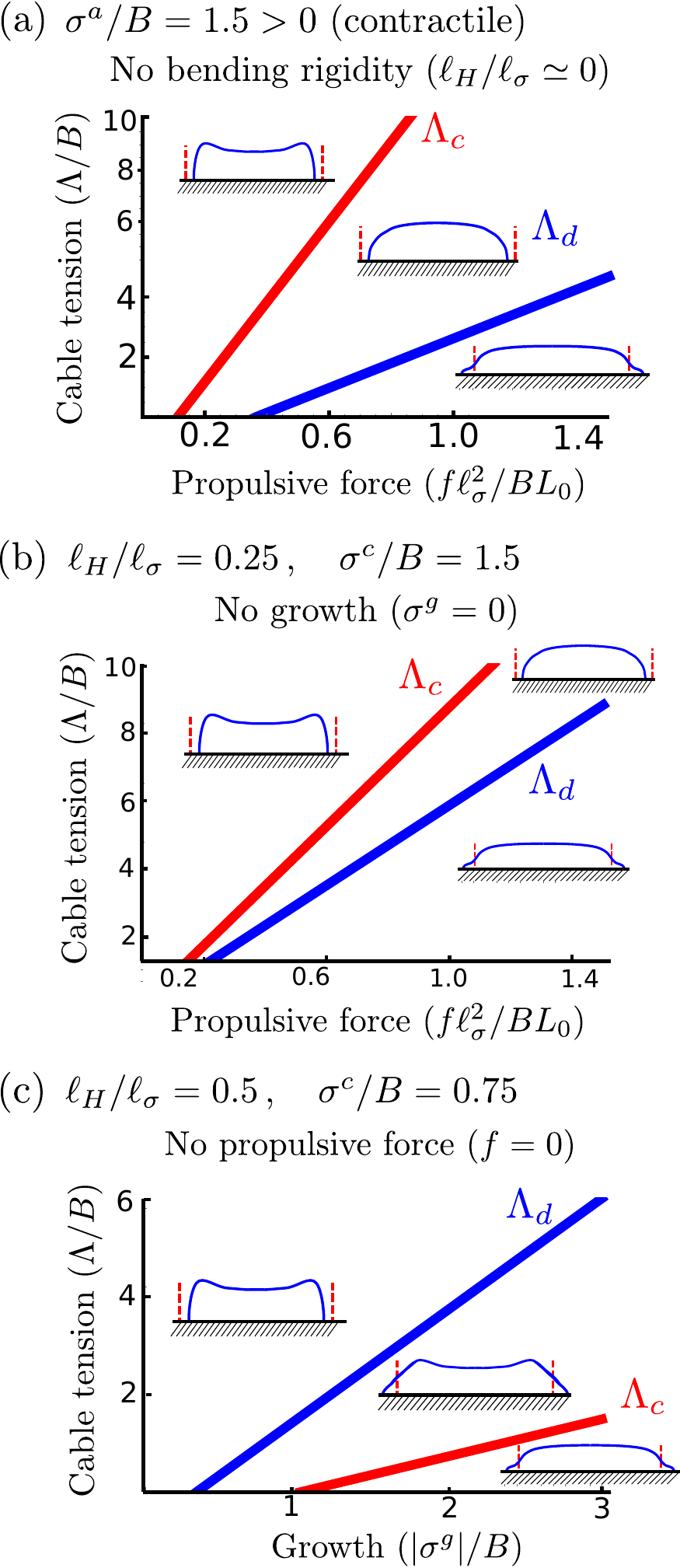}
	\caption{Morphological phase diagram showing the curvature transition at $\Lambda_c$ (red line) and the size-change transition at $\Lambda_d$ (blue line). In all three plots, we fix $\ell_p/\ell_{\sigma}=0.7$ and $\ell_\sigma/L_0\ll 1$. Note that for finite $\ell_H$, the bulk active stresses, $\sigma^c$ and $\sigma^g$ enter independently and the width of the region between $\Lambda_d$ and $\Lambda_c$ is controlled by the relative size of $\sigma^c$, $|\sigma^g|$ and $f$. In all three figures the total bulk active stress $\sigma_a$ is taken to be dominantly contractile, hence, there is a minimum extensile stress (either from $|\sigma^g|$ or $f$) required for $\Lambda_{c,d}$ to exist. In (a), the bending rigidity of the tissue is neglected ($\ell_H/\ell_{\sigma}\to 0$) and $\sigma_a>0$, with both $\sigma^g$ and $\sigma^c$ finite. Note that in this limit, bulk active stresses only appear together in the additive combination $\sigma_a$, unlike the finite $\ell_H$ case. In (b) $\sigma_a>0$, but with $\sigma^g=0$. Turning on a small yet finite $\ell_H\neq 0$, we obtain a qualitatively similar phase diagram as in (a), i.e., with $\Lambda_d<\Lambda_c$, so tissue contraction occurs \emph{prior} to curvature change upon increasing $\Lambda$. Including a small $\sigma^g<0$ only moves the $\Lambda_{c,d}$ intercepts to the left as the bulk active stress $\sigma_a$ decreases. In (c), the behavior is shown as a function of $|\sigma^g|$ for $f=0$ and a higher $\ell_H$. We find that in this case $\Lambda_c\neq\Lambda_d$ even for $f=0$. Additionally, different from (a) and (b), now $\Lambda_d>\Lambda_c$ and hence the tissue contracts in-plane \emph{after} changing its apical curvature.}
	\label{fig:phasedgm}
\end{figure}

Distinct from the curvature change, the displacement of the tissue boundary changes sign at a different threshold for $f\neq 0$, given by
 \begin{equation}
	\Lambda_d=fL_d-\sigma_a\;.\label{eq:Lambdad}
\end{equation}
To see this we set $x=L_0$ in Eq.~\ref{eq:usoln} to obtain
\begin{gather}
	u_x(L_0)=\dfrac{1}{Y_sL_\sigma}\tanh\left(\dfrac{L_0}{L_\sigma}\right)\left[-\sigma_a+(fL_d-\Lambda)\right]\;,\label{eq:uL}\\
	L_d=\dfrac{L_\sigma}{\tanh(L_0/L_\sigma)}\left[1+\dfrac{\ell_\sigma^2}{\ell_p^2-L_\sigma^2}\left(1-\dfrac{\ell_p\tanh(L_0/L_\sigma)}{L_\sigma\tanh(L_0/\ell_p)}\right)\right]\;.
\end{gather}
Here $L_d$ is the length scale that captures the influence of the propulsive force on the tissue displacement. For a large tissue, we have
\begin{equation}
	L_d\approx L_\sigma-\dfrac{\ell^2_\sigma}{(\ell_p+L_\sigma)}\;,\quad L_\sigma,\ell_p\ll L_0\;,\label{eq:Ld}
\end{equation}
which is positive as $L_\sigma>\ell_\sigma$. Unlike $L_c$, $L_d$ is independent of the tissue size for a large tissue, irrespective of the ratio $\ell_p/L_\sigma$ and is primarily controlled by the stress penetration depth $L_\sigma$. This highlights the distinction between the force transmission mechanisms that control curvature and shape of the tissue versus its size and adhesive properties. It is useful to contrast this with Ref.~\cite{perez2019active}, where a size dependent dewetting transition was observed in an epithelial tissue modeled as an active fluid, which albeit different, is nonetheless similar~\footnote{Note that by replacing displacements with velocities, the planar expansion-contraction change of the elastic tissue exactly corresponds to the dewetting transition of its fluid counterpart.} to our elastic model. The main distinction lies in the strength of cell-substrate adhesions ($Y_s$), which in Ref.~\cite{perez2019active} is considered negligible, resulting in $L_\sigma\gg L_0$, whereas, we work in the strongly adhered limit with $L_\sigma\ll L_0$. As a consequence, our elastic expansion-contraction transtion is size independent. On the other hand, for weak substrate adhesion, we can replace $L_\sigma$ by $L_0$ in Eq.~\ref{eq:Ld}, thereby recovering the size dependence seen Ref.~\cite{perez2019active}, albeit now in an elastic model. We also find qualitative agreement  with the measured stress profiles~\cite{perez2019active} in this parameter regime where the stress is dominated by bulk contractility and peaked in the interior. From Eq.~\ref{eq:uL}, we easily find that $u_x(L_0)$ changes sign at the value $\Lambda=\Lambda_d$ given in Eq.~(\ref{eq:Lambdad}).
Hence, as we change $\Lambda$, we go through a tissue size transition, where for
\begin{itemize}
	\item $\Lambda>\Lambda_d$: the tissue is globally contracted.
	\item $\Lambda<\Lambda_d$: the tissue is globally extended.
\end{itemize}
Importantly, when $\ell_p\ll L_\sigma$, $\Lambda_d>\Lambda_c$, while for $\ell_p\gtrsim L_\sigma$, $\Lambda_d<\Lambda_c$ and $\Lambda_c$ is then size dependent\footnote{Of course, as before, when the bulk active stresses are dominantly contractile ($\sigma_a>0$), either transition exists only for a sufficiently large propulsive force.}. As $\Lambda_c\neq\Lambda_d$ when $f\neq 0$, we find that our model predicts four different morphological states for the tissue as sketched in Fig.~\ref{fig:shapes}. An illustrative morphological ``phase diagram'' is shown in Fig.~\ref{fig:phasedgm}a for $\sigma_a>0$, in the $\Lambda$-$f$ plane. Changes in the stress profile from being peaked near the tissue center to being peaked near the boundary with a nonmonotonic spatial profile have been reported previously in epithelial monolayers \cite{blanch2017effective} and our results are in qualitative agreement.
Note that, just like the stress profile, the tissue displacement is also nonmonotonic in general (see Fig.~\ref{fig:nonmon}). So while the edge of the tissue contracts from its rest length when $\Lambda>\Lambda_d$, the center of the tissue can be locally extended due to the stress being more extensile there and vice-versa. As a result, while our simple characterization in terms of just $H(0)$ and $u_x(L_0)$ is easy to understand, the full tissue shape and stress profile can be accessed in experiments through imaging and traction force microscopy allowing for more stringent tests of our theory.

\subsection{Role of apical bending rigidity}
Until now, we focused on the minimal model where the bending rigidity of the apical surface $\kappa$ was neglected in favour of its surface tension $\gamma$. This allowed us to take $\ell_H=\sqrt{\kappa/\gamma}\to 0$ and slave $H$ to the stress profile (Eq.~\ref{eq:slave}). Reintroducting a finite but small $\ell_H\ll \ell_\sigma,\ell_p$ does not change the above results, but larger values of $\ell_H$ do affect the tissue morphology and the transitions in qualitative ways. While the curvature is once again slaved to the total stress  in the bulk of the tissue, this is no longer the case on scales $\sim\ell_H$ near the boundary. Using the fact that the contractile ring generates a boundary torque that enforces $H(L_0)\propto(\sigma^c+3\Lambda)>0$, we see that, close to the tissue boundary, the variation of the curvature on a length scale $\sim\ell_H$ provides an additional effective source of localized stress in Eq.~\ref{eq:sigmaeqn1d} through the passive term $(Bh_0/12)H$. As a result, we find an extra positive contribution $\sim(\sigma^c+3\Lambda)$ to the force balance equation localized over a region $\ell_H$ from the boundary. This additional contribution enters at the same level as the polarization term, but with the opposite sign. Hence, we can easily extend our previous results by viewing the effect of a finite $\ell_H$ as providing an additional  contractile force near the edge spread out over a region of size $\ell_H$, akin to an effective \emph{negative} propulsive force. This is a direct consequence of apico-basal polarity in the tissue that permits active torques on the apical surface. An immediate implication is that, in the absence of extensile forces, such as arising from growth or polarized cell motility ($f=\sigma^g=0$), neither a curvature nor a planar size-changing transition can occur in the tissue, even for finite $\ell_H$. Alternately, even in the absence of polarized motility ($f=0$), for a finite $\ell_H$ and $\sigma_a<0$, the curvature change and expansion-contraction transitions now don't coincide. The various states and transition boundaries, including a finite $\ell_H$ as well, are plotted in the morphological phase diagram shown in Fig.~\ref{fig:phasedgm}.

\section{Conclusion}
\label{sec:conclusion}
In this paper, by using a lubrication approximation, we have developed a simple 2D elastic model for epithelial tissues strongly adhered to a flat rigid substrate. Crucially, we incorporate both apicobasal polarity in the tissue and the local variation of cellular thickness, allowing us to address the consequences of active stresses on tissue shape. The morphology of a resting epithelium is decided by a competition between bulk and boundary active stresses in conjunction with the elasticity of the tissue and substrate adhesion. We distinguish two kinds of transitions, one concerning the curvature of the apical surface and another for the in-plane size change of the tissue. The basic physics underlying these shape changes is transparent: extensile stresses (like positive internal pressure) cause the apical surface to be ``dome-shaped'' and locally expand the tissue, while contractile stresses (like negative pressure) do the opposite, as expected.

Within a minimal model that neglects the bending rigidity of the apical surface, the curvature $H$ can be slaved entirely to the total stress in the tissue. In this limit, the transition of either tissue shape or size are decided by a balance of bulk active stresses including contractility and growth $\sim\sigma^c+\sigma^g$, the actomyosin cable tension $\sim\Lambda$ and the net stress $\sim -fL_{c,d}$ arising from cellular motility at the leading edge (remember that $-f\b{p}$ is the force exerted \emph{by} the tissue). The length scale $L_{c,d}$ over which propulsive forces are transmitted is decided by the elastic parameters and differs in general for the two transitions. In particular, $L_c$ can be size dependent, while $L_d$ is not in general for a large tissue. Including a finite bending rigidity has a similar effect as an effective \emph{negative} propulsive force as a result of a cumulative transmission of active torques generated by differential apicobasal contractility and the boundary actomyosin cable. Although we only consider homogeneous bulk active stresses, an edge localized spatial profile of either growth or contractility would also have the same effect as the propulsive force, only with the overall sign determined by the stress contribution being mostly contractile or extensile.

In the past few years, there has been a growing understanding on the mechanical basis of tissue morphogenesis in controlled settings, such as in organoids \cite{karzbrun2018human}. Recent \emph{in-vitro} experiments \cite{lei2013modulation,ishida2014epithelial} demonstrate that epithelial tissues can initiate lumen formation through a folding transition when exposed to a bath of extracellular matrix (ECM). It is conceivable that such a shape change is triggered by a mechanism involving competing bulk and boundary active stresses as in our model. There is some evidence that the recruitment of ECM components such as laminin can potentially reinforce actomyosin contractility around the edge of a tissue \cite{colognato1999laminin}, thereby increasing $\Lambda$ in our model. This would provide a useful experimental knob to traverse the morphological phase diagram in Fig.~\ref{fig:phasedgm}. A useful test would be to measure the stress profile along with the tissue curvature and compare against our continuum results, as has been done previously for expanding monolayers viewed as an active fluid \cite{blanch2017effective}, though without reference to apical curvature.

Stress profiles in epithelial monolayers reported previously \cite{blanch2017effective,perez2019active} agree qualitatively with our elastic model, suggesting the fluid versus elastic dichotomy isn't easily discriminated by stresses alone. More recently, active torques arising from a polarized distribution of actomyosin have been experimentally quantified in freely suspended epithelia \cite{fouchard2019curling}, highlighting the importance of such torques in bending tissues. While apical curvature provides a distinct morphological phenotype, it is largely unexplored, and we hope our work encourages further investigation and experimental probes of tissue curvature. Our work provides insight into the routes by which active forces can shape planar stationary epithelia, and extending these results to curved surfaces and time-dependent nonlinear phenomena are the next immediate challenges.

\section{Conflicts of interest}
There are no conflicts to declare.

\section{Acknowledgements}
We would like to thank Eyal Karzbrun, Sebastian Streichan and Boris Shraiman for insightful discussions. This work is primarily supported by the National Science Foundation (NSF) through the Materials Science and Engineering Center at UC Santa Barbara, DMR-1720256 (iSuperSeed), with additional support from NSF grants DMR-1609208 (MCM, BL, FS and SS) and PHY-1748958 (KITP). SS is supported by the Harvard Society of Fellows. MJB, BL, FS and SS would like to acknowledge the hospitality of KITP, where some of this work was done.


\section{\LARGE Appendices}
\bigskip

\begin{appendices}
\section{Parametrizing the strain tensor}
\label{app:strain}
\begin{figure}[]
	\centering
	\includegraphics[width=0.45\textwidth]{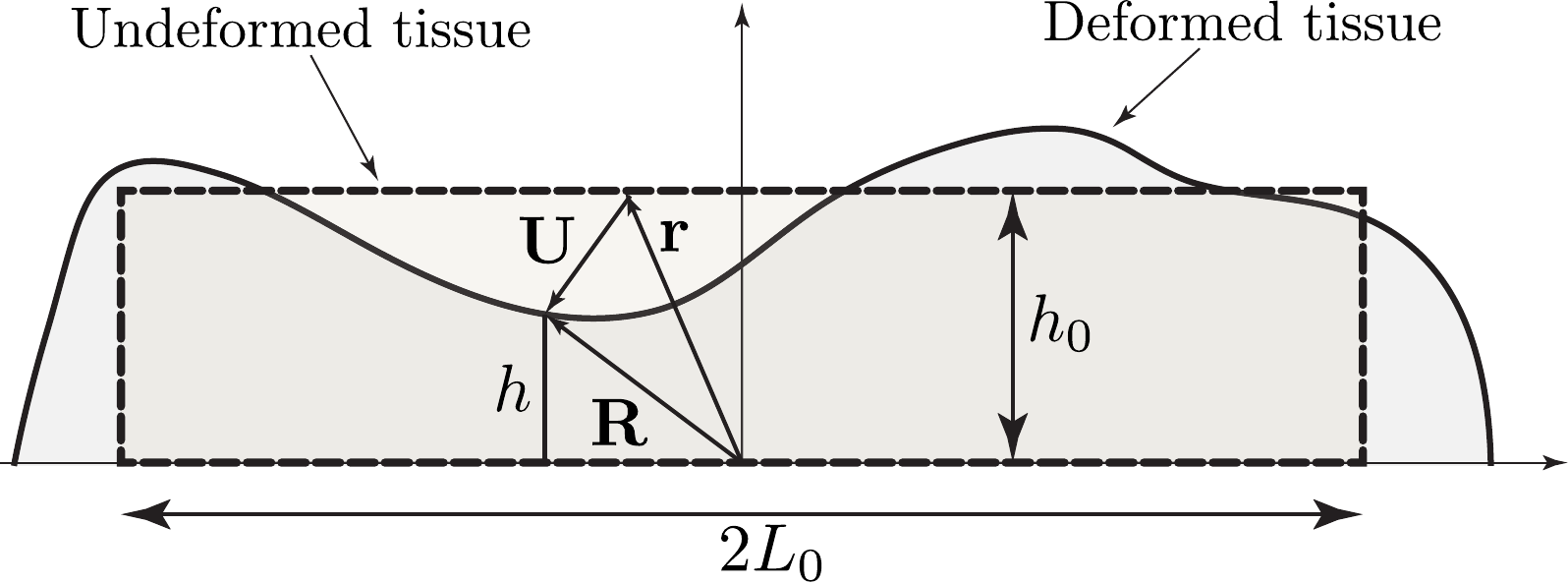}
	\caption{Sketch of the coordinate system used to parametrize the tissue showing both the undeformed and deformed geometries.}
	\label{fig:cartoon2}
\end{figure}
	In this Appendix, we parametrize the tissue deformation in terms of in-plane displacements ($\b{u}$) and a height field for the tissue thickness ($h$). This is done by enforcing $\eps_{iz}=0$ as stated in the main text. Writing the 3D position of any point in tissue as $\b{R}$, we have the identity
\begin{equation}
	\b{R}(x,y,z)=\b{R}_0(x,y)+\int_0^z\dd z'\partial_{z'}\b{R}(x,y,z')\;,\label{eq:R}
\end{equation}
	where $\b{R}_0\equiv\b{R}(z=0)$ and $\b{R}_0\cdot\hat{\b{z}}=0$ as the basal surface is attached to a planar substrate. In the undeformed tissue, $\partial_z\b{R}=\hat{\b{z}}$ and it continues to specify the normal to a local $x-y$ section of the deformed tissue as well. Writing $\partial_z\b{R}=\hat{\b{z}}+\b{w}$, where $\b{w}$ is a small deflection, we set $2\eps_{iz}=\partial_iR_z+\partial_zR_i=0$ to linear order in $\b{w}$. Consistency requires that
\begin{equation}
	w_i=-\int_0^z\dd z' \partial_iw_z(z')\;,\quad i=x,y\;,
\end{equation}
	while $w_z$ is not constrained as of yet. As the basal surface is planar, $\partial_z\b{R}(z=0)=\hat{\b{z}}$, hence $w_z(z=0)=0$. The thickness of the tissue being small, we Taylor expand $w_z$ as a function of $z$ and retain the lowest order term, which is
\begin{equation}
	w_z=\dfrac{z}{h_0}W(x,y)\;.
\end{equation}
	This simple linear interpolation is a convenient ansatz for the 3D deformation of the tissue and is the most dominant term for a thin tissue. The function $W(x,y)$, as we will see, is related to the local thickness of the tissue. Using this parametrization in Eq.~\ref{eq:R}, we obtain $\b{R}=\b{r}+\b{U}$, where $\b{r}=(x,y,z)$ is the undeformed material coordinate and the 3D displacement $\b{U}$ is
\begin{equation}
	U_i=u_i-\dfrac{z^3}{6h_0}\partial_iW\;,\quad U_z=\dfrac{z^2}{2h_0}W\;,
\end{equation}
	having introduced the in-plane 2D displacement $\b{u}$ such that $\b{R}_0=(x+u_x,y+u_y,0)$. The deformed thickness of the tissue is obtained from $\hat{\b{z}}\cdot\b{R}(z=h_0)=h$, which relates $W$ and $h$ as
\begin{equation}
	W=2\left(\dfrac{h-h_0}{h_0}\right)\;.
\end{equation}
Hence, $W$ is exactly the strain in the $z$-direction. This completes our parametrization of the 3D displacement, from which it is trivial to obtain the strain tensor quoted in the main text (Eq.~\ref{eq:epsij})

\section{Stationary solution in a circular geometry}
\label{app:circle}
Here we consider an axisymmetric tissue in a circular geometry of radius $R$. Once again defining $H=\del^2h$ as the mean curvature and $\sigma=(\sigma_{rr}+\sigma_{\phi\phi})/2$ as the average normal stress, we can rewrite Eqs.~\ref{eq:ueqn} and~\ref{eq:heqn} at steady state in terms of $H$ and $\sigma$ as follows
\begin{align}
	\ell_{\sigma}^2\del^2\sigma&=\sigma-\sigma_a+\left(\dfrac{1-\nu}{2}\right)\ell_{\sigma}^2\del^2\sigma_a+(1+\nu)\dfrac{Bh_0}{24}H\nonumber\\
	&\qquad\qquad\qquad-\left(\dfrac{1+\nu}{2}\right)f\ell_{\sigma}^2\bm\del\cdot\b{p}\;,\\
	\ell_{H}^2\del^2H&=H+\dfrac{h_0}{\gamma(1+\nu)}\left[\nu\del^2\sigma^c+\del^2\sigma-\left(\dfrac{1-\nu}{2}\right)\del^2\sigma^g\right]\;.
\end{align}
	We have similarly defined the average contractile and growth induced stresses as $\sigma^{c,g}=\tr(\vec{\sigma}^{c,g})/2$ along with the average active stress $\sigma_a=\sigma^c+\sigma^g$. Using circular polar coordinates, we only have a radial dependence for $H$ and $\sigma$ in the axisymmetric case. Slaving $H$ to $\sigma$ in the $\ell_{H}\to0$ limit as before and taking $\sigma^{a}$ to be spatially constant for simplicity, we obtain
\begin{equation}
	L_{\sigma}^2\del^2\sigma=\sigma-\sigma_a-\ell_{\sigma}^2\left(\dfrac{1+\nu}{2}\right)\bm\del\cdot\b{p}\;,\label{eq:sig1}
\end{equation}
	with the same $L_{\sigma}$ as before. Notice that this equation has the same form as the 1D model discussed in the main text (Eq.~\ref{eq:sig0}). Similarly, writing $\b{u}=u_r(r)\hat{\b{r}}$, the displacement satisfies the following simple equation,
\begin{equation}
	L_{\sigma}^2\left(\del^2 u_r-\dfrac{u_r}{r^2}\right)=u_r+\dfrac{f}{Y_s}p_r\left[1+\dfrac{L_\sigma^2-\ell_\sigma^2}{\ell_p^2}\right]\;,
\end{equation}
where we have used $\b{p}=p_r(r)\hat{\b{r}}$.
Solving Eq.~\ref{eq:peqn} in the circular domain for the polarization profile along with $p_r(R)=1$ we obtain,
\begin{equation}
	p_r(r)=\dfrac{I_1(r/\ell_p)}{I_1(R/\ell_p)}\;,
\end{equation}
	where $I_{\alpha}(x)$ is the modified Bessel function of the first kind. Note that while $H$ is slaved to $-\del^2\sigma$, the stress boundary condition involves only the radial component, $\sigma_{rr}(R)=-\tilde{\Lambda}/R$ ($C=1/R$) and $\sigma_{r\phi}=0$ everywhere due to axisymmetry. Using the individual stress components and their respective boundary conditions, we obtain
\begin{align}
	\sigma_{rr}(r)&=\sigma_a-\left(\dfrac{\tilde{\Lambda}}{R}+\sigma_a\right)\dfrac{F(r/L_\sigma)}{F(R/L_\sigma)}\nonumber\\
	&+f\dfrac{\ell_\sigma^2\ell_p}{(\ell_p^2-L_\sigma^2)}\dfrac{F(R/\ell_p)}{I_1(R/\ell_p)}\left[\dfrac{F(r/\ell_p)}{F(R/\ell_p)}-\dfrac{F(r/L_\sigma)}{F(R/L_\sigma)}\right]\;,\\
	\sigma_{\phi\phi}(r)&=\sigma_a-\left(\dfrac{\tilde{\Lambda}}{R}+\sigma_a\right)\dfrac{G(r/L_\sigma)}{F(R/L_\sigma)}\nonumber\\
	&-f\dfrac{\ell_\sigma^2\ell_p}{(\ell_p^2-L_\sigma^2)}\dfrac{F(R/\ell_p)}{I_1(R/\ell_p)}\left[\dfrac{G(r/\ell_p)}{F(R/\ell_p)}-\dfrac{G(r/L_\sigma)}{F(R/L_\sigma)}\right]\;,\\
	\sigma(r)&=\sigma_a-\dfrac{(1+\nu)}{2}\left\{\left(\dfrac{\tilde{\Lambda}}{R}+\sigma_a\right)\dfrac{I_0(r/L_\sigma)}{F(R/L_\sigma)}\right.\nonumber\\
	&\left.-f\dfrac{\ell_\sigma^2\ell_p}{(\ell_p^2-L_\sigma^2)}\dfrac{F(R/\ell_p)}{I_1(R/\ell_p)}\left[\dfrac{I_0(r/\ell_p)}{F(R/\ell_p)}-\dfrac{I_0(r/L_\sigma)}{F(R/L_\sigma)}\right]\right\}\;.
\end{align}
Here we have defined two auxiliary functions
\begin{align}
	F(x)&=I_0(x)-\dfrac{(1-\nu)}{x}I_1(x)\;,\\
	G(x)&=\nu I_0(x)+\dfrac{(1-\nu)}{x}I_1(x)\;.
\end{align}
Note that $F$ and $G$ have simple asymptotics, $F(x)\approx e^x/\sqrt{2\pi x}$ and $G(x)\approx \nu e^{x}/\sqrt{2\pi x}$ as $x\to\infty$.
The displacement and curvature are similarly obtained to be
\begin{align}
	u_r(r)&=-\dfrac{(\sigma_a+\tilde{\Lambda}/R)}{Y_sL_\sigma}\dfrac{I_1(r/L_\sigma)}{F(R/L_\sigma)}+\dfrac{f}{Y_s}\left\{\vphantom{\left[\dfrac{L_\sigma^2}{\ell_p^2}\right]}\dfrac{I_1(r/\ell_p)}{I_1(R/\ell_p)}\right.\nonumber\\
	&\left.+\dfrac{\ell_\sigma^2}{(\ell_p^2-L_\sigma^2)}\dfrac{F(R/\ell_p)}{I_1(R/\ell_p)}\left[\dfrac{I_1(r/\ell_p)}{F(R/\ell_p)}-\dfrac{\ell_p}{L_\sigma}\dfrac{I_1(r/L_\sigma)}{F(R/L_\sigma)}\right]\right\}\;,\\
	H(r)&=\dfrac{h_0}{2\gamma}\left\{\dfrac{(\sigma_a+\tilde{\Lambda}/R)}{L_\sigma^2}\dfrac{I_0(r/L_\sigma)}{F(R/L_\sigma)}\vphantom{\left[\dfrac{I_0(R/L_\sigma)}{I_1(R/L_\sigma)}\left(\dfrac{\ell_p}{L_\sigma}\right)^2\right]}\right.\nonumber\\
	&\left.+f\dfrac{\ell_\sigma^2}{\ell_p(\ell_p^2-L_\sigma^2)}\dfrac{F(R/\ell_p)}{I_1(R/\ell_p)}\left[\dfrac{I_0(r/\ell_p)}{F(R/\ell_p)}-\left(\dfrac{\ell_p}{L_\sigma}\right)^2\dfrac{I_0(r/L_\sigma)}{F(R/L_\sigma)}\right]\right\}\;.
\end{align}
Proceeding as in the 1D model and writing $\Lambda=\tilde{\Lambda}/R$, we have
\begin{equation}
	H(0)=\dfrac{h_0}{2\gamma L_\sigma^2 F(R/L_\sigma)}\left[\sigma_a+\Lambda-f L_c\right]\;
\end{equation}
which changes sign at $\Lambda_c=fL_c-\sigma_a$ just as in the main text (Eq.~\ref{eq:Lambdac2}). The length scale
\begin{equation}
	L_c=\dfrac{\ell_\sigma^2}{\ell_p(\ell_p^2-L_\sigma^2)}\left[\dfrac{L_\sigma^2F(R/L_\sigma)-\ell_p^2F(R/\ell_p)}{I_1(R/\ell_p)}\right]\;,
\end{equation}
has identical scaling behaviour with respect to $\ell_p$ and $\ell_\sigma$ in a large tissue ($R\gg L_\sigma,\ell_p$) as in the simple 1D model. Similarly, $u_r(R)=(I_1(R/L_\sigma)/Y_sL_\sigma F(R/L_\sigma))[fL_d-\Lambda-\sigma_a]$ changes sign at $\Lambda_d=fL_d-\sigma_a$, with
\begin{equation}
	L_d=\dfrac{L_\sigma F(R/L_\sigma)}{I_1(R/L_\sigma)}\left[1+\dfrac{\ell_\sigma^2}{\ell_p^2-L_\sigma^2}\left(1-\dfrac{\ell_pF(R/\ell_p)I_1(R/L_\sigma)}{L_\sigma F(R/L_\sigma)I_1(R/\ell_p)}\right)\right]\;,
\end{equation}
which in a large tissue scales the same way as in the 1D model (Eq.~\ref{eq:Ld}). Hence the simple 1D model captures all the same physics, along with the qualitative spatial profiles as in the more involved 2D axisymmetric circular tissue.

\end{appendices}
\balance

\begin{mcitethebibliography}{57}
\providecommand*{\natexlab}[1]{#1}
\providecommand*{\mciteSetBstSublistMode}[1]{}
\providecommand*{\mciteSetBstMaxWidthForm}[2]{}
\providecommand*{\mciteBstWouldAddEndPuncttrue}
  {\def\EndOfBibitem{\unskip.}}
\providecommand*{\mciteBstWouldAddEndPunctfalse}
  {\let\EndOfBibitem\relax}
\providecommand*{\mciteSetBstMidEndSepPunct}[3]{}
\providecommand*{\mciteSetBstSublistLabelBeginEnd}[3]{}
\providecommand*{\EndOfBibitem}{}
\mciteSetBstSublistMode{f}
\mciteSetBstMaxWidthForm{subitem}
{(\emph{\alph{mcitesubitemcount}})}
\mciteSetBstSublistLabelBeginEnd{\mcitemaxwidthsubitemform\space}
{\relax}{\relax}

\bibitem[Lecuit and Le~Goff(2007)]{lecuit2007orchestrating}
T.~Lecuit and L.~Le~Goff, \emph{Nature}, 2007, \textbf{450}, 189\relax
\mciteBstWouldAddEndPuncttrue
\mciteSetBstMidEndSepPunct{\mcitedefaultmidpunct}
{\mcitedefaultendpunct}{\mcitedefaultseppunct}\relax
\EndOfBibitem
\bibitem[Nelson and Gleghorn(2012)]{nelson2012sculpting}
C.~M. Nelson and J.~P. Gleghorn, \emph{Annual review of biomedical
  engineering}, 2012, \textbf{14}, 129--154\relax
\mciteBstWouldAddEndPuncttrue
\mciteSetBstMidEndSepPunct{\mcitedefaultmidpunct}
{\mcitedefaultendpunct}{\mcitedefaultseppunct}\relax
\EndOfBibitem
\bibitem[Lecuit and Lenne(2007)]{lecuit2007cell}
T.~Lecuit and P.-F. Lenne, \emph{Nature reviews Molecular cell biology}, 2007,
  \textbf{8}, 633\relax
\mciteBstWouldAddEndPuncttrue
\mciteSetBstMidEndSepPunct{\mcitedefaultmidpunct}
{\mcitedefaultendpunct}{\mcitedefaultseppunct}\relax
\EndOfBibitem
\bibitem[Montell(2008)]{montell2008morphogenetic}
D.~J. Montell, \emph{Science}, 2008, \textbf{322}, 1502--1505\relax
\mciteBstWouldAddEndPuncttrue
\mciteSetBstMidEndSepPunct{\mcitedefaultmidpunct}
{\mcitedefaultendpunct}{\mcitedefaultseppunct}\relax
\EndOfBibitem
\bibitem[Trepat and Sahai(2018)]{trepat2018mesoscale}
X.~Trepat and E.~Sahai, \emph{Nature Physics}, 2018, \textbf{14},
  671--682\relax
\mciteBstWouldAddEndPuncttrue
\mciteSetBstMidEndSepPunct{\mcitedefaultmidpunct}
{\mcitedefaultendpunct}{\mcitedefaultseppunct}\relax
\EndOfBibitem
\bibitem[Xi \emph{et~al.}(2019)Xi, Saw, Delacour, Lim, and
  Ladoux]{xi2019material}
W.~Xi, T.~B. Saw, D.~Delacour, C.~T. Lim and B.~Ladoux, \emph{Nature Reviews
  Materials}, 2019, \textbf{4}, 23--44\relax
\mciteBstWouldAddEndPuncttrue
\mciteSetBstMidEndSepPunct{\mcitedefaultmidpunct}
{\mcitedefaultendpunct}{\mcitedefaultseppunct}\relax
\EndOfBibitem
\bibitem[Villar \emph{et~al.}(2013)Villar, Graham, and
  Bayley]{villar2013tissue}
G.~Villar, A.~D. Graham and H.~Bayley, \emph{Science}, 2013, \textbf{340},
  48--52\relax
\mciteBstWouldAddEndPuncttrue
\mciteSetBstMidEndSepPunct{\mcitedefaultmidpunct}
{\mcitedefaultendpunct}{\mcitedefaultseppunct}\relax
\EndOfBibitem
\bibitem[Ideses \emph{et~al.}(2018)Ideses, Erukhimovitch, Brand, Jourdain,
  Hernandez, Gabinet, Safran, Kruse, and
  Bernheim-Groswasser]{ideses2018spontaneous}
Y.~Ideses, V.~Erukhimovitch, R.~Brand, D.~Jourdain, J.~S. Hernandez,
  U.~Gabinet, S.~Safran, K.~Kruse and A.~Bernheim-Groswasser, \emph{Nature
  communications}, 2018, \textbf{9}, 2461\relax
\mciteBstWouldAddEndPuncttrue
\mciteSetBstMidEndSepPunct{\mcitedefaultmidpunct}
{\mcitedefaultendpunct}{\mcitedefaultseppunct}\relax
\EndOfBibitem
\bibitem[Senoussi \emph{et~al.}(2019)Senoussi, Kashida, Voituriez, Galas,
  Maitra, and Est{\'e}vez-Torres]{senoussi2019tunable}
A.~Senoussi, S.~Kashida, R.~Voituriez, J.-C. Galas, A.~Maitra and
  A.~Est{\'e}vez-Torres, \emph{Proceedings of the National Academy of
  Sciences}, 2019, \textbf{116}, 22464--22470\relax
\mciteBstWouldAddEndPuncttrue
\mciteSetBstMidEndSepPunct{\mcitedefaultmidpunct}
{\mcitedefaultendpunct}{\mcitedefaultseppunct}\relax
\EndOfBibitem
\bibitem[Morley \emph{et~al.}(2019)Morley, Ellison, Bhattacharjee, O'Bryan,
  Zhang, Smith, Kabb, Sebastian, Moore,
  Schulze,\emph{et~al.}]{morley2019quantitative}
C.~D. Morley, S.~T. Ellison, T.~Bhattacharjee, C.~S. O'Bryan, Y.~Zhang, K.~F.
  Smith, C.~P. Kabb, M.~Sebastian, G.~L. Moore, K.~D. Schulze \emph{et~al.},
  \emph{Nature communications}, 2019, \textbf{10}, 1--9\relax
\mciteBstWouldAddEndPuncttrue
\mciteSetBstMidEndSepPunct{\mcitedefaultmidpunct}
{\mcitedefaultendpunct}{\mcitedefaultseppunct}\relax
\EndOfBibitem
\bibitem[Gonzalez-Rodriguez \emph{et~al.}(2012)Gonzalez-Rodriguez, Guevorkian,
  Douezan, and Brochard-Wyart]{gonzalez2012soft}
D.~Gonzalez-Rodriguez, K.~Guevorkian, S.~Douezan and F.~Brochard-Wyart,
  \emph{Science}, 2012, \textbf{338}, 910--917\relax
\mciteBstWouldAddEndPuncttrue
\mciteSetBstMidEndSepPunct{\mcitedefaultmidpunct}
{\mcitedefaultendpunct}{\mcitedefaultseppunct}\relax
\EndOfBibitem
\bibitem[Nelson(2016)]{nelson2016buckling}
C.~M. Nelson, \emph{Journal of biomechanical engineering}, 2016, \textbf{138},
  021005\relax
\mciteBstWouldAddEndPuncttrue
\mciteSetBstMidEndSepPunct{\mcitedefaultmidpunct}
{\mcitedefaultendpunct}{\mcitedefaultseppunct}\relax
\EndOfBibitem
\bibitem[Park \emph{et~al.}(2015)Park, Kim, Bi, Mitchel, Qazvini, Tantisira,
  Park, McGill, Kim, Gweon,\emph{et~al.}]{park2015unjamming}
J.-A. Park, J.~H. Kim, D.~Bi, J.~A. Mitchel, N.~T. Qazvini, K.~Tantisira, C.~Y.
  Park, M.~McGill, S.-H. Kim, B.~Gweon \emph{et~al.}, \emph{Nature materials},
  2015, \textbf{14}, 1040\relax
\mciteBstWouldAddEndPuncttrue
\mciteSetBstMidEndSepPunct{\mcitedefaultmidpunct}
{\mcitedefaultendpunct}{\mcitedefaultseppunct}\relax
\EndOfBibitem
\bibitem[Noll \emph{et~al.}(2017)Noll, Mani, Heemskerk, Streichan, and
  Shraiman]{noll2017active}
N.~Noll, M.~Mani, I.~Heemskerk, S.~J. Streichan and B.~I. Shraiman,
  \emph{Nature physics}, 2017, \textbf{13}, 1221\relax
\mciteBstWouldAddEndPuncttrue
\mciteSetBstMidEndSepPunct{\mcitedefaultmidpunct}
{\mcitedefaultendpunct}{\mcitedefaultseppunct}\relax
\EndOfBibitem
\bibitem[Latorre \emph{et~al.}(2018)Latorre, Kale, Casares,
  G{\'o}mez-Gonz{\'a}lez, Uroz, Valon, Nair, Garreta, Montserrat, del
  Campo,\emph{et~al.}]{latorre2018active}
E.~Latorre, S.~Kale, L.~Casares, M.~G{\'o}mez-Gonz{\'a}lez, M.~Uroz, L.~Valon,
  R.~V. Nair, E.~Garreta, N.~Montserrat, A.~del Campo \emph{et~al.},
  \emph{Nature}, 2018, \textbf{563}, 203\relax
\mciteBstWouldAddEndPuncttrue
\mciteSetBstMidEndSepPunct{\mcitedefaultmidpunct}
{\mcitedefaultendpunct}{\mcitedefaultseppunct}\relax
\EndOfBibitem
\bibitem[Armon \emph{et~al.}(2018)Armon, Bull, Aranda-Diaz, and
  Prakash]{armon2018ultrafast}
S.~Armon, M.~S. Bull, A.~Aranda-Diaz and M.~Prakash, \emph{Proceedings of the
  National Academy of Sciences}, 2018, \textbf{115}, E10333--E10341\relax
\mciteBstWouldAddEndPuncttrue
\mciteSetBstMidEndSepPunct{\mcitedefaultmidpunct}
{\mcitedefaultendpunct}{\mcitedefaultseppunct}\relax
\EndOfBibitem
\bibitem[Martin \emph{et~al.}(2009)Martin, Kaschube, and
  Wieschaus]{martin2009pulsed}
A.~C. Martin, M.~Kaschube and E.~F. Wieschaus, \emph{Nature}, 2009,
  \textbf{457}, 495\relax
\mciteBstWouldAddEndPuncttrue
\mciteSetBstMidEndSepPunct{\mcitedefaultmidpunct}
{\mcitedefaultendpunct}{\mcitedefaultseppunct}\relax
\EndOfBibitem
\bibitem[Kim \emph{et~al.}(2013)Kim, Varner, and Nelson]{kim2013apical}
H.~Y. Kim, V.~D. Varner and C.~M. Nelson, \emph{Development}, 2013,
  \textbf{140}, 3146--3155\relax
\mciteBstWouldAddEndPuncttrue
\mciteSetBstMidEndSepPunct{\mcitedefaultmidpunct}
{\mcitedefaultendpunct}{\mcitedefaultseppunct}\relax
\EndOfBibitem
\bibitem[Dervaux and Amar(2008)]{dervaux2008morphogenesis}
J.~Dervaux and M.~B. Amar, \emph{Physical review letters}, 2008, \textbf{101},
  068101\relax
\mciteBstWouldAddEndPuncttrue
\mciteSetBstMidEndSepPunct{\mcitedefaultmidpunct}
{\mcitedefaultendpunct}{\mcitedefaultseppunct}\relax
\EndOfBibitem
\bibitem[Shyer \emph{et~al.}(2013)Shyer, Tallinen, Nerurkar, Wei, Gil, Kaplan,
  Tabin, and Mahadevan]{shyer2013villification}
A.~E. Shyer, T.~Tallinen, N.~L. Nerurkar, Z.~Wei, E.~S. Gil, D.~L. Kaplan,
  C.~J. Tabin and L.~Mahadevan, \emph{Science}, 2013, \textbf{342},
  212--218\relax
\mciteBstWouldAddEndPuncttrue
\mciteSetBstMidEndSepPunct{\mcitedefaultmidpunct}
{\mcitedefaultendpunct}{\mcitedefaultseppunct}\relax
\EndOfBibitem
\bibitem[Liang and Mahadevan(2011)]{liang2011growth}
H.~Liang and L.~Mahadevan, \emph{Proceedings of the National Academy of
  Sciences}, 2011, \textbf{108}, 5516--5521\relax
\mciteBstWouldAddEndPuncttrue
\mciteSetBstMidEndSepPunct{\mcitedefaultmidpunct}
{\mcitedefaultendpunct}{\mcitedefaultseppunct}\relax
\EndOfBibitem
\bibitem[Armon \emph{et~al.}(2011)Armon, Efrati, Kupferman, and
  Sharon]{armon2011geometry}
S.~Armon, E.~Efrati, R.~Kupferman and E.~Sharon, \emph{Science}, 2011,
  \textbf{333}, 1726--1730\relax
\mciteBstWouldAddEndPuncttrue
\mciteSetBstMidEndSepPunct{\mcitedefaultmidpunct}
{\mcitedefaultendpunct}{\mcitedefaultseppunct}\relax
\EndOfBibitem
\bibitem[Hannezo \emph{et~al.}(2014)Hannezo, Prost, and
  Joanny]{hannezo2014theory}
E.~Hannezo, J.~Prost and J.-F. Joanny, \emph{Proceedings of the National
  Academy of Sciences}, 2014, \textbf{111}, 27--32\relax
\mciteBstWouldAddEndPuncttrue
\mciteSetBstMidEndSepPunct{\mcitedefaultmidpunct}
{\mcitedefaultendpunct}{\mcitedefaultseppunct}\relax
\EndOfBibitem
\bibitem[Hannezo \emph{et~al.}(2011)Hannezo, Prost, and
  Joanny]{hannezo2011instabilities}
E.~Hannezo, J.~Prost and J.-F. Joanny, \emph{Physical Review Letters}, 2011,
  \textbf{107}, 078104\relax
\mciteBstWouldAddEndPuncttrue
\mciteSetBstMidEndSepPunct{\mcitedefaultmidpunct}
{\mcitedefaultendpunct}{\mcitedefaultseppunct}\relax
\EndOfBibitem
\bibitem[Maitra \emph{et~al.}(2014)Maitra, Srivastava, Rao, and
  Ramaswamy]{maitra2014activating}
A.~Maitra, P.~Srivastava, M.~Rao and S.~Ramaswamy, \emph{Physical review
  letters}, 2014, \textbf{112}, 258101\relax
\mciteBstWouldAddEndPuncttrue
\mciteSetBstMidEndSepPunct{\mcitedefaultmidpunct}
{\mcitedefaultendpunct}{\mcitedefaultseppunct}\relax
\EndOfBibitem
\bibitem[Berthoumieux \emph{et~al.}(2014)Berthoumieux, Ma{\^\i}tre, Heisenberg,
  Paluch, J{\"u}licher, and Salbreux]{berthoumieux2014active}
H.~Berthoumieux, J.-L. Ma{\^\i}tre, C.-P. Heisenberg, E.~K. Paluch,
  F.~J{\"u}licher and G.~Salbreux, \emph{New Journal of Physics}, 2014,
  \textbf{16}, 065005\relax
\mciteBstWouldAddEndPuncttrue
\mciteSetBstMidEndSepPunct{\mcitedefaultmidpunct}
{\mcitedefaultendpunct}{\mcitedefaultseppunct}\relax
\EndOfBibitem
\bibitem[Murisic \emph{et~al.}(2015)Murisic, Hakim, Kevrekidis, Shvartsman, and
  Audoly]{murisic2015discrete}
N.~Murisic, V.~Hakim, I.~G. Kevrekidis, S.~Y. Shvartsman and B.~Audoly,
  \emph{Biophysical journal}, 2015, \textbf{109}, 154--163\relax
\mciteBstWouldAddEndPuncttrue
\mciteSetBstMidEndSepPunct{\mcitedefaultmidpunct}
{\mcitedefaultendpunct}{\mcitedefaultseppunct}\relax
\EndOfBibitem
\bibitem[Salbreux and J{\"u}licher(2017)]{salbreux2017mechanics}
G.~Salbreux and F.~J{\"u}licher, \emph{Physical Review E}, 2017, \textbf{96},
  032404\relax
\mciteBstWouldAddEndPuncttrue
\mciteSetBstMidEndSepPunct{\mcitedefaultmidpunct}
{\mcitedefaultendpunct}{\mcitedefaultseppunct}\relax
\EndOfBibitem
\bibitem[Mietke \emph{et~al.}(2019)Mietke, J{\"u}licher, and
  Sbalzarini]{mietke2019self}
A.~Mietke, F.~J{\"u}licher and I.~F. Sbalzarini, \emph{Proceedings of the
  National Academy of Sciences}, 2019, \textbf{116}, 29--34\relax
\mciteBstWouldAddEndPuncttrue
\mciteSetBstMidEndSepPunct{\mcitedefaultmidpunct}
{\mcitedefaultendpunct}{\mcitedefaultseppunct}\relax
\EndOfBibitem
\bibitem[Krajnc and Ziherl(2015)]{krajnc2015theory}
M.~Krajnc and P.~Ziherl, \emph{Physical Review E}, 2015, \textbf{92},
  052713\relax
\mciteBstWouldAddEndPuncttrue
\mciteSetBstMidEndSepPunct{\mcitedefaultmidpunct}
{\mcitedefaultendpunct}{\mcitedefaultseppunct}\relax
\EndOfBibitem
\bibitem[Morris and Rao(2019)]{morris2019active}
R.~G. Morris and M.~Rao, \emph{Physical Review E}, 2019, \textbf{100},
  022413\relax
\mciteBstWouldAddEndPuncttrue
\mciteSetBstMidEndSepPunct{\mcitedefaultmidpunct}
{\mcitedefaultendpunct}{\mcitedefaultseppunct}\relax
\EndOfBibitem
\bibitem[Serra-Picamal \emph{et~al.}(2012)Serra-Picamal, Conte, Vincent, Anon,
  Tambe, Bazellieres, Butler, Fredberg, and Trepat]{serra2012mechanical}
X.~Serra-Picamal, V.~Conte, R.~Vincent, E.~Anon, D.~T. Tambe, E.~Bazellieres,
  J.~P. Butler, J.~J. Fredberg and X.~Trepat, \emph{Nature Physics}, 2012,
  \textbf{8}, 628\relax
\mciteBstWouldAddEndPuncttrue
\mciteSetBstMidEndSepPunct{\mcitedefaultmidpunct}
{\mcitedefaultendpunct}{\mcitedefaultseppunct}\relax
\EndOfBibitem
\bibitem[K{\"o}pf and Pismen(2013)]{kopf2013continuum}
M.~H. K{\"o}pf and L.~M. Pismen, \emph{Soft Matter}, 2013, \textbf{9},
  3727--3734\relax
\mciteBstWouldAddEndPuncttrue
\mciteSetBstMidEndSepPunct{\mcitedefaultmidpunct}
{\mcitedefaultendpunct}{\mcitedefaultseppunct}\relax
\EndOfBibitem
\bibitem[Brugu{\'e}s \emph{et~al.}(2014)Brugu{\'e}s, Anon, Conte, Veldhuis,
  Gupta, Colombelli, Mu{\~n}oz, Brodland, Ladoux, and
  Trepat]{brugues2014forces}
A.~Brugu{\'e}s, E.~Anon, V.~Conte, J.~H. Veldhuis, M.~Gupta, J.~Colombelli,
  J.~J. Mu{\~n}oz, G.~W. Brodland, B.~Ladoux and X.~Trepat, \emph{Nature
  physics}, 2014, \textbf{10}, 683\relax
\mciteBstWouldAddEndPuncttrue
\mciteSetBstMidEndSepPunct{\mcitedefaultmidpunct}
{\mcitedefaultendpunct}{\mcitedefaultseppunct}\relax
\EndOfBibitem
\bibitem[Banerjee \emph{et~al.}(2015)Banerjee, Utuje, and
  Marchetti]{banerjee2015propagating}
S.~Banerjee, K.~J. Utuje and M.~C. Marchetti, \emph{Physical review letters},
  2015, \textbf{114}, 228101\relax
\mciteBstWouldAddEndPuncttrue
\mciteSetBstMidEndSepPunct{\mcitedefaultmidpunct}
{\mcitedefaultendpunct}{\mcitedefaultseppunct}\relax
\EndOfBibitem
\bibitem[Blanch-Mercader \emph{et~al.}(2017)Blanch-Mercader, Vincent,
  Bazelli{\`e}res, Serra-Picamal, Trepat, and Casademunt]{blanch2017effective}
C.~Blanch-Mercader, R.~Vincent, E.~Bazelli{\`e}res, X.~Serra-Picamal, X.~Trepat
  and J.~Casademunt, \emph{Soft Matter}, 2017, \textbf{13}, 1235--1243\relax
\mciteBstWouldAddEndPuncttrue
\mciteSetBstMidEndSepPunct{\mcitedefaultmidpunct}
{\mcitedefaultendpunct}{\mcitedefaultseppunct}\relax
\EndOfBibitem
\bibitem[Douezan and Brochard-Wyart(2012)]{douezan2012dewetting}
S.~Douezan and F.~Brochard-Wyart, \emph{The European Physical Journal E}, 2012,
  \textbf{35}, 34\relax
\mciteBstWouldAddEndPuncttrue
\mciteSetBstMidEndSepPunct{\mcitedefaultmidpunct}
{\mcitedefaultendpunct}{\mcitedefaultseppunct}\relax
\EndOfBibitem
\bibitem[Ravasio \emph{et~al.}(2015)Ravasio, Le, Saw, Tarle, Ong, Bertocchi,
  M{\`e}ge, Lim, Gov, and Ladoux]{ravasio2015regulation}
A.~Ravasio, A.~P. Le, T.~B. Saw, V.~Tarle, H.~T. Ong, C.~Bertocchi, R.-M.
  M{\`e}ge, C.~T. Lim, N.~S. Gov and B.~Ladoux, \emph{Integrative Biology},
  2015, \textbf{7}, 1228--1241\relax
\mciteBstWouldAddEndPuncttrue
\mciteSetBstMidEndSepPunct{\mcitedefaultmidpunct}
{\mcitedefaultendpunct}{\mcitedefaultseppunct}\relax
\EndOfBibitem
\bibitem[P{\'e}rez-Gonz{\'a}lez \emph{et~al.}(2019)P{\'e}rez-Gonz{\'a}lez,
  Alert, Blanch-Mercader, G{\'o}mez-Gonz{\'a}lez, Kolodziej, Bazellieres,
  Casademunt, and Trepat]{perez2019active}
C.~P{\'e}rez-Gonz{\'a}lez, R.~Alert, C.~Blanch-Mercader,
  M.~G{\'o}mez-Gonz{\'a}lez, T.~Kolodziej, E.~Bazellieres, J.~Casademunt and
  X.~Trepat, \emph{Nature Physics}, 2019, \textbf{15}, 79\relax
\mciteBstWouldAddEndPuncttrue
\mciteSetBstMidEndSepPunct{\mcitedefaultmidpunct}
{\mcitedefaultendpunct}{\mcitedefaultseppunct}\relax
\EndOfBibitem
\bibitem[Joanny and Ramaswamy(2012)]{joanny2012drop}
J.-F. Joanny and S.~Ramaswamy, \emph{Journal of fluid mechanics}, 2012,
  \textbf{705}, 46--57\relax
\mciteBstWouldAddEndPuncttrue
\mciteSetBstMidEndSepPunct{\mcitedefaultmidpunct}
{\mcitedefaultendpunct}{\mcitedefaultseppunct}\relax
\EndOfBibitem
\bibitem[Ciarlet(1997)]{ciarlet1997mathematical}
P.~G. Ciarlet, \emph{Mathematical Elasticity: Volume II: Theory of Plates},
  Elsevier, 1997, vol.~27\relax
\mciteBstWouldAddEndPuncttrue
\mciteSetBstMidEndSepPunct{\mcitedefaultmidpunct}
{\mcitedefaultendpunct}{\mcitedefaultseppunct}\relax
\EndOfBibitem
\bibitem[Banerjee and Marchetti(2019)]{banerjee2019continuum}
S.~Banerjee and M.~C. Marchetti, \emph{Cell Migrations: Causes and Functions},
  Springer, 2019, pp. 45--66\relax
\mciteBstWouldAddEndPuncttrue
\mciteSetBstMidEndSepPunct{\mcitedefaultmidpunct}
{\mcitedefaultendpunct}{\mcitedefaultseppunct}\relax
\EndOfBibitem
\bibitem[Marchetti \emph{et~al.}(2013)Marchetti, Joanny, Ramaswamy, Liverpool,
  Prost, Rao, and Simha]{marchetti2013hydrodynamics}
M.~C. Marchetti, J.-F. Joanny, S.~Ramaswamy, T.~B. Liverpool, J.~Prost, M.~Rao
  and R.~A. Simha, \emph{Reviews of Modern Physics}, 2013, \textbf{85},
  1143\relax
\mciteBstWouldAddEndPuncttrue
\mciteSetBstMidEndSepPunct{\mcitedefaultmidpunct}
{\mcitedefaultendpunct}{\mcitedefaultseppunct}\relax
\EndOfBibitem
\bibitem[Prost \emph{et~al.}(2015)Prost, J{\"u}licher, and
  Joanny]{prost2015active}
J.~Prost, F.~J{\"u}licher and J.-F. Joanny, \emph{Nature physics}, 2015,
  \textbf{11}, 111--117\relax
\mciteBstWouldAddEndPuncttrue
\mciteSetBstMidEndSepPunct{\mcitedefaultmidpunct}
{\mcitedefaultendpunct}{\mcitedefaultseppunct}\relax
\EndOfBibitem
\bibitem[Banerjee \emph{et~al.}(2019)Banerjee, Sarkar, Toner, and
  Basu]{banerjee2019rolled}
T.~Banerjee, N.~Sarkar, J.~Toner and A.~Basu, \emph{Physical Review Letters},
  2019, \textbf{122}, 218002\relax
\mciteBstWouldAddEndPuncttrue
\mciteSetBstMidEndSepPunct{\mcitedefaultmidpunct}
{\mcitedefaultendpunct}{\mcitedefaultseppunct}\relax
\EndOfBibitem
\bibitem[Ramaswamy \emph{et~al.}(2000)Ramaswamy, Toner, and
  Prost]{ramaswamy2000nonequilibrium}
S.~Ramaswamy, J.~Toner and J.~Prost, \emph{Physical review letters}, 2000,
  \textbf{84}, 3494\relax
\mciteBstWouldAddEndPuncttrue
\mciteSetBstMidEndSepPunct{\mcitedefaultmidpunct}
{\mcitedefaultendpunct}{\mcitedefaultseppunct}\relax
\EndOfBibitem
\bibitem[Hutson \emph{et~al.}(2003)Hutson, Tokutake, Chang, Bloor, Venakides,
  Kiehart, and Edwards]{hutson2003forces}
M.~S. Hutson, Y.~Tokutake, M.-S. Chang, J.~W. Bloor, S.~Venakides, D.~P.
  Kiehart and G.~S. Edwards, \emph{Science}, 2003, \textbf{300}, 145--149\relax
\mciteBstWouldAddEndPuncttrue
\mciteSetBstMidEndSepPunct{\mcitedefaultmidpunct}
{\mcitedefaultendpunct}{\mcitedefaultseppunct}\relax
\EndOfBibitem
\bibitem[Rodriguez-Diaz \emph{et~al.}(2008)Rodriguez-Diaz, Toyama, Abravanel,
  Wiemann, Wells, Tulu, Edwards, and Kiehart]{rodriguez2008actomyosin}
A.~Rodriguez-Diaz, Y.~Toyama, D.~L. Abravanel, J.~M. Wiemann, A.~R. Wells,
  U.~S. Tulu, G.~S. Edwards and D.~P. Kiehart, \emph{HFSP journal}, 2008,
  \textbf{2}, 220--237\relax
\mciteBstWouldAddEndPuncttrue
\mciteSetBstMidEndSepPunct{\mcitedefaultmidpunct}
{\mcitedefaultendpunct}{\mcitedefaultseppunct}\relax
\EndOfBibitem
\bibitem[Kiehart(1999)]{kiehart1999wound}
D.~P. Kiehart, \emph{Current biology}, 1999, \textbf{9}, R602--R605\relax
\mciteBstWouldAddEndPuncttrue
\mciteSetBstMidEndSepPunct{\mcitedefaultmidpunct}
{\mcitedefaultendpunct}{\mcitedefaultseppunct}\relax
\EndOfBibitem
\bibitem[Jacinto \emph{et~al.}(2001)Jacinto, Martinez-Arias, and
  Martin]{jacinto2001mechanisms}
A.~Jacinto, A.~Martinez-Arias and P.~Martin, \emph{Nature cell biology}, 2001,
  \textbf{3}, E117\relax
\mciteBstWouldAddEndPuncttrue
\mciteSetBstMidEndSepPunct{\mcitedefaultmidpunct}
{\mcitedefaultendpunct}{\mcitedefaultseppunct}\relax
\EndOfBibitem
\bibitem[N{\"a}rv{\"a} \emph{et~al.}(2017)N{\"a}rv{\"a}, Stubb, Guzm{\'a}n,
  Blomqvist, Balboa, Lerche, Saari, Otonkoski, and Ivaska]{narva2017strong}
E.~N{\"a}rv{\"a}, A.~Stubb, C.~Guzm{\'a}n, M.~Blomqvist, D.~Balboa, M.~Lerche,
  M.~Saari, T.~Otonkoski and J.~Ivaska, \emph{Stem cell reports}, 2017,
  \textbf{9}, 67--76\relax
\mciteBstWouldAddEndPuncttrue
\mciteSetBstMidEndSepPunct{\mcitedefaultmidpunct}
{\mcitedefaultendpunct}{\mcitedefaultseppunct}\relax
\EndOfBibitem
\bibitem[Trepat \emph{et~al.}(2009)Trepat, Wasserman, Angelini, Millet, Weitz,
  Butler, and Fredberg]{trepat2009physical}
X.~Trepat, M.~R. Wasserman, T.~E. Angelini, E.~Millet, D.~A. Weitz, J.~P.
  Butler and J.~J. Fredberg, \emph{Nature physics}, 2009, \textbf{5}, 426\relax
\mciteBstWouldAddEndPuncttrue
\mciteSetBstMidEndSepPunct{\mcitedefaultmidpunct}
{\mcitedefaultendpunct}{\mcitedefaultseppunct}\relax
\EndOfBibitem
\bibitem[Karzbrun \emph{et~al.}(2018)Karzbrun, Kshirsagar, Cohen, Hanna, and
  Reiner]{karzbrun2018human}
E.~Karzbrun, A.~Kshirsagar, S.~R. Cohen, J.~H. Hanna and O.~Reiner,
  \emph{Nature physics}, 2018, \textbf{14}, 515\relax
\mciteBstWouldAddEndPuncttrue
\mciteSetBstMidEndSepPunct{\mcitedefaultmidpunct}
{\mcitedefaultendpunct}{\mcitedefaultseppunct}\relax
\EndOfBibitem
\bibitem[Lei \emph{et~al.}(2013)Lei, Zouani, Rami, Chanseau, and
  Durrieu]{lei2013modulation}
Y.~Lei, O.~F. Zouani, L.~Rami, C.~Chanseau and M.-C. Durrieu, \emph{Small},
  2013, \textbf{9}, 1086--1095\relax
\mciteBstWouldAddEndPuncttrue
\mciteSetBstMidEndSepPunct{\mcitedefaultmidpunct}
{\mcitedefaultendpunct}{\mcitedefaultseppunct}\relax
\EndOfBibitem
\bibitem[Ishida \emph{et~al.}(2014)Ishida, Tanaka, Yamaguchi, Ogata, Mizutani,
  Kawabata, and Haga]{ishida2014epithelial}
S.~Ishida, R.~Tanaka, N.~Yamaguchi, G.~Ogata, T.~Mizutani, K.~Kawabata and
  H.~Haga, \emph{PloS one}, 2014, \textbf{9}, e99655\relax
\mciteBstWouldAddEndPuncttrue
\mciteSetBstMidEndSepPunct{\mcitedefaultmidpunct}
{\mcitedefaultendpunct}{\mcitedefaultseppunct}\relax
\EndOfBibitem
\bibitem[Colognato \emph{et~al.}(1999)Colognato, Winkelmann, and
  Yurchenco]{colognato1999laminin}
H.~Colognato, D.~A. Winkelmann and P.~D. Yurchenco, \emph{The Journal of cell
  biology}, 1999, \textbf{145}, 619--631\relax
\mciteBstWouldAddEndPuncttrue
\mciteSetBstMidEndSepPunct{\mcitedefaultmidpunct}
{\mcitedefaultendpunct}{\mcitedefaultseppunct}\relax
\EndOfBibitem
\bibitem[Fouchard \emph{et~al.}(2019)Fouchard, Wyatt, Proag, Lisica,
  Khalilgharibi, Recho, Suzanne, Kabla, and Charras]{fouchard2019curling}
J.~Fouchard, T.~Wyatt, A.~Proag, A.~Lisica, N.~Khalilgharibi, P.~Recho,
  M.~Suzanne, A.~Kabla and G.~Charras, \emph{bioRxiv}, 2019,  806455\relax
\mciteBstWouldAddEndPuncttrue
\mciteSetBstMidEndSepPunct{\mcitedefaultmidpunct}
{\mcitedefaultendpunct}{\mcitedefaultseppunct}\relax
\EndOfBibitem
\end{mcitethebibliography}

\providecommand*{\mcitethebibliography}{\thebibliography}
\csname @ifundefined\endcsname{endmcitethebibliography}
{\let\endmcitethebibliography\endthebibliography}{}

\end{document}